\documentclass[11pt,preprint2]{aastex} 

\slugcomment{Not to appear in Nonlearned J., 45.}

\shorttitle{$BVRI$ Surface Photometry of Isolated Spiral Galaxies.}

\shortauthors{Hern\'andez-Toledo et al.}

\begin{document}

\title{$BVRI$ Surface Photometry of Isolated Spiral Galaxies\thanks{Based on data 
obtained at the 0.84m and 1.5m telescopes of the Observatorio Astron\'omico Nacional, San Pedro M\'artir, 
Baja California, M\'exico, operated by the Instituto de Astronom\'{\i}a, Universidad Nacional Aut\'onoma de M\'exico.}}
\author{H. M. Hern\'andez-Toledo\altaffilmark{1} J. Zendejas-Dom\'{\i}nguez\altaffilmark{2} 
and 
V. Avila-Reese\altaffilmark{3}}
\affil{Instituto de Astronom\'{\i}a, Universidad Nacional Aut\'onoma de M\'exico, A.P. 70-264, 04510 
M\'exico D. F., M\'exico}

\altaffiltext{1}{E-mail: hector@astroscu.unam.mx}
\altaffiltext{2}{E-mail: zendejas@astroscu.unam.mx}
\altaffiltext{3}{E-mail: avila@astroscu.unam.mx}

\begin{abstract}
A release of multicolor broad band ($BVRI$) photometry for a subsample of 44 isolated 
spirals drawn from the {\it Catalogue of Isolated Galaxies} (CIG) is presented. Total 
magnitudes and colors at various circular apertures, as well as some global 
structural/morphological parameters are estimated. Morphology is reevaluated 
through optical and sharp/filtered $R$ band images, ($B-I$) color index maps, 
and archive near--IR $JHK$ images from the 
Two--Micron Survey. The $CAS$ structural parameters (Concentration, Asymmetry, 
and Clumpiness) were calculated from the images in each one of the bands.
The fraction of galaxies with well identified optical/near--IR bars (SB) is 63\%, 
while a 17\% more shows evidence of weak or suspected bars (SAB). The sample 
average value of the maximum bar ellipticity is $\epsilon_{\rm max}\approx 0.4$. 
Half of the galaxies in the sample shows rings. We identify two candidates for 
isolated galaxies with disturbed morphology. The structural $CAS$ parameters change
with the observed band, and the tendencies they follow with the
morphological type and global color are more evident in the redder bands.
In any band, the major difference between our isolated spirals and a sample
of interacting spirals is revealed in the $A-S$ plane. A deep and uniformly 
observed sample of isolated galaxies is intended for various purposes including 
(i) comparative studies of environmental effects, (ii) confronting model predictions 
of galaxy evolution and (iii) evaluating the change of galaxy properties with redshift.

\end{abstract}


\section{Introduction} \label{S1}

The concept of a "field" population of galaxies as distinct from the group/cluster 
populations has existed since the earliest days of extragalactic astronomy (Hubble 
1936) and it is used recurrently in studies aimed to explore the effects of
large--scale environment on galaxy properties. However, the definition of 
``field'' is fuzzy. The distribution of galaxies in space is actually strongly 
clustered and a large fraction of them is prone to form gravitationally-bound 
multiple systems, from very populated clusters to loose groups, the majority 
being in normal groups (Tully 1987). {\it Isolation} is an important requirement 
beyond the concept of ``field'' galaxies. A galaxy is isolated if it has not suffered
any interaction with another normal galaxy or with a group environment over a
Hubble time or at least since approximately one half of its mass was assembled. 
This makes the observational finding and study of isolated galaxies important because,
among other reasons, (i) they can be used as comparison objects in studies of 
the environmental effects on galaxies belonging to groups and clusters, and 
(ii) they are ideal for confronting with theoretical and model 
predictions of galaxy evolution.

The fact that properties of galaxies change with environment has been known for a long time.
The main observable dependencies with environment --from cluster centers to the
sparse ``field''-- are seen for the morphological mix, the global colors, and 
the specific star formation (SF) rate; for recent reviews see Park et al. (2007),
Avila-Reese et  al. (2005), and the references therein. These properties are inferred 
mainly from  photometric observations. In fact, to study observationally the effects of high--density 
environments on the galaxy properties, a good understanding of the isolated, 
non--perturbed, galaxies is necessary.   These galaxies are also required as 
a control sample for studying interacting galaxies. For example, in 
Hern\'andez-Toledo et al. (2005; see also Conselice 2003) we have compared
the photometric concentration, asymmetry and clumpiness parameters
of local interacting and ``field'' disk galaxies with the aim to establish a 
relatively easy way for identifying interacting disk galaxies in high--redshfit samples.

Among isolated galaxies, the observational study of disk galaxies is of 
special interest because, on one hand, they are expected to be more drastically 
affected by environmental and interaction effects; on the other hand, according 
to the current paradigm of cosmic structure formation, the formation of disks 
inside hierarchically growing Cold Dark Matter (CDM) haloes is a generic process. 
A large amount of work has been done in modeling the evolution of (isolated) 
disk galaxies. Well observed samples of isolated disk galaxies are hence important 
in order to confront with model predictions (e.g., Avila--Reese \& Firmani 2000; 
Cole et al. 2000; Boissier \& Prantzos 2000; Firmani \& Avila--Reese 2000; Yang, 
Mo \& van den Bosch 2003; Zavala et al. 2003; Pizagno et al. 2005; Dutton et al. 2006; 
Kassin, de Jong \& Weiner 2006; Gnedin et al. 2006).

 A uniformly selected and observed sample of isolated disk galaxies is also crucial for 
studying intrinsic secular processes able to affect the structure, morphology and
dynamics of galaxies, for instance, the formation and evolution of bars, circular
rings, lopsidedness, and bulges. However, chances are that isolated galaxies may 
show evidence of disturbances not associated with intrinsic processes, which opens
the necessity to explore other alternatives.
On this line, cosmological numerical simulations within the CDM model show that inside 
the galaxy--size haloes there survives a large population of subhaloes (Klypin et al.
1999; Moore et al. 1999), but reionization and feedback could inhibit the 
formation of luminous (satellite) galaxies inside most of these subhaloes 
(e.g., Bullock, Kravtsov \& Weinberg 2000; Benson et al. 2002). The
subhalos and the associated gas clouds could produce signs of distortion 
on isolated galaxies (Threntam, Moller \& Ramirez-Ruiz 2001; Pisano, Wilcots 
\& Liu 2002). If these signs are clear, then observations in radio could reveal the 
presence of 21--cm hydrogen line emission associated to a pure gas companion  
galaxy. But if such an emission is absent, there is still the possibility that 
the 'perturber' is just a dark matter (sub)halo without any baryonic component 
(Threntam et al. 2001).
Recently, Karachentsev, Karachentseva \& Huchtmeier (2006), by analyzing a large
sample of isolated galaxies, have reported the finding of four such possible cases.

Homogeneous observational data samples of isolated 
disk galaxies are crucial for obtaining transparent scaling relationships and
correlations that can be appropriately confronted with model predictions (see,
e.g. Zavala et al. 2003).  In recent years, some groups 
have been working on the compilation and observation of such samples (e.g., Pisano 
et al. 2002; Allam et al. 2005; Koopmann \& Kenney 2006).
For galaxies in voids, which are the most likely to be isolated, see 
Rojas et al. (2004,2005).
It is worth to mention that the important requirements for all these samples
are: well defined and strong isolation criteria, a uniform-quality data 
acquisition in several wavelengths, and completeness when possible.

We have carried out optical CCD photometry for a representative set  
of galaxies in the northern Catalogue of Isolated Galaxies (hereafter CIG, 
Karachentseva 1973). This is one of the best defined and most complete catalogues 
of isolated galaxies. The aim of this paper is to present a global $BVRI$
photometric and morphological analysis for a subsample of 44 spiral galaxies 
from the CIG catalog. It is strongly emphasized that all 
the observations were done with the same CCD detector. After applying uniform reduction 
and analysis procedures, an homogeneous set of photometric and morphological data is 
guaranteed. CIG galaxies cover a wide range in luminosities, surface brightnesses, 
morphological types and colors. Their relative simplicity and closeness, as compared 
with galaxies in other environments, offer a unique opportunity to have a more detailed 
and less confused interpretation of their structural, photometric and morphological 
properties.  
 
The outline of the paper is as follows. Section \ref{DataSample} summarizes the 
selection criteria applied to the isolated galaxy sample that are relevant to 
our photometric study, and describes the observations and reduction techniques 
used here. Section \ref{Magnitudes} presents a
comparison of our estimated total magnitudes against those in the literature. 
In Section \ref{Morphology}, we discuss the observed morphology 
based on mosaic $R-$band and $R-$band Sharp/filtered images, $(B-I)$ color index 
maps, and composed near--infrared (NIR) $JHK$ images extracted from the Two--Micron 
Survey archives. Emphasis is put on the presence of disturbed morphology. Section 
\ref{PhysMorphology} presents our estimates of the optical ($BVRI$) and 
NIR $JK-$band concentration, asymmetry, and clumpiness ($CAS$) structural parameters. 
In Section \ref{S4} we explore and discuss some basic correlations among the 
$BVRI-JK$ photometric and structural parameters in this 
sample that could be useful for comparative studies involving galaxies in other 
environments. Section \ref{S5} provides a summary of the paper. 
Finally, an Appendix is devoted to the presentation of $BVRI$ magnitudes at 
two other concentric circular apertures.    
 
\section{The Data Sample} \label{DataSample}

\subsection{Isolated Spiral Galaxies from Kara\-chentseva Catalogue.}

We have carried out an observational program at the Observatorio Astron\'omico Nacional
at San Pedro M\'artir (OAN-SPM), Baja California, M\'exico, devoted to obtain 
uniform CCD photometric data for one of the most complete and homogeneous samples of 
isolated galaxies currently available, the CIG catalogue of Karachentseva (1973). 
This sample amounts to more than 1050 galaxies in the northern hemisphere. 
The CCD $BVRI$ images in the Johnson-Cousins system
were obtained with a Site1 detector attached to the 1.5m and 0.84m telescopes at OAN-SPM 
covering an area of about $4.3\arcmin \times 4.3\arcmin$ and $7.2\arcmin \times 7.2\arcmin$, 
a typical seeing of 1.7 arcsec and a scale of 0.51$\arcsec$/pixel and 0.85$\arcsec$/pixel respectively. 

The original number of galaxies in three observing runs amounts to 52 galaxies. From these, 6 galaxies 
obtained under bad observing conditions and 2 ellipticals   
were eliminated, yielding a final 
sample of 44 isolated spiral galaxies of the present study. We applied no special strategy in selecting this current 
subset. Availability of 
observing time and weather conditions were the main factors constraining 
the number of observed galaxies. Some aspects of the selection criteria for the CIG sample that are most relevant 
to the present and further photometric analyses are stated here. 
 
The isolated galaxies in the CIG sample were selected from a visual search of the Palomar Sky Survey. The catalogue samples
the sky north of $\delta \geq -3\degr$. The vast majority of  objects is found in high Galactic latitude regions ($b \geq  20\degr$) 
and as a sample, it is reasonably complete ($\sim 90$\%) in the magnitude range $13.5 \leq m_{zw} \leq 15.7$ 
(Hernandez-Toledo et al. 1999).
The selection criteria used in assembling the CIG can be expressed by the following relations:
 
\begin{eqnarray}
x_{1i} > 20 a_{i} \cr 
0.25 a_{1} < a_{i} < 4 a_{1}, 
\end{eqnarray}
where$x_{1i}$  is the apparent separation between the candidate isolated galaxy  
of apparent diameter $a_{1}$ and any other neighbor galaxy of apparent diameter 
$a_{i}$.  Under these criteria, any other galaxy of comparable size ($a_{i} = a_{1}$) 
should be at a distance of at least 20 times its diameter (projected on 
the sky) from the isolated galaxy. 

 Assuming a typical galaxy diameter D $\sim$ 20 kpc and a peculiar velocity 
relative to the Hubble flow V $\sim 150$ km s$^{-1}$ (Rivolo \& Yahil 1981), the time 
required for an intruder galaxy to traverse 20 diameters is $\sim 2\times 10^{9}$ 
yr. This is a first-order estimate of the time since the last equal-size 
galaxy-galaxy interaction for a CIG system and it suggests that CIG galaxies are 
reasonably isolated. It is therefore expected that only the intrinsic properties of the 
individual galaxies in the 
CIG  should influence the observed photometric and morphological properties.

Karachentseva (1973) included other isolation criteria (coded as 1 and 2 in 
her original catalogue) depending on whether or not other galaxies, within a 
factor of 4 in size, were found  near a 20 diameter boundary or even if a galaxy 
is definitely not isolated according to the CIG criteria.  Less isolated 
galaxies account for less than $5\%$ of our CIG sample and have been excluded 
from the present study.

\subsection{Data Reduction}

A journal of  the photometric observations is given in Table 1. Column (1) gives the original 
catalogue number, Columns (2)-(9) give the number of frames per filter, the integration time (in seconds), and seeing 
conditions (in arcsec).

\placetable{tbl-1}

Table 2 reports some relevant information on the observed isolated galaxies obtained from the literature. 
Column (1) is the CIG catalogue number, Column (2) reports other identifications, Column (3) the apparent total $B$ 
magnitude from the Lyon Extragalactic Database (LEDA), Column (4) the Hubble Type from LEDA, Column (5) the apparent 
total $B$ magnitude from the Nasa Extragalactic Database (NED),  Column (6) the radial velocity in km s$^{-1}$ 
corrected for Virgocentric infall from LEDA.

\placetable{tbl-2}

Images were debiased, trimmed, and flat-fielded using standard IRAF\footnote{The 
IRAF package is written and supported by the IRAF programming group at the National 
Optical Astronomy Observatories (NOAO) in Tucson, Arizona. NOAO is operated by the 
Association of Universities for Research in Astronomy (AURA), Inc. under cooperative 
agreement with the National Science Foundation (NSF).} procedures. First, the bias 
level of the CCD was subtracted from all exposures. A run of 10 bias images 
was obtained per night, and those were combined into a single bias frame which 
was then applied to the object frames. The images were flat-fielded using sky 
flats taken in each filter at the beginning and/or end of each night.  
 
Photometric calibration was achieved by nightly observations of standard stars of 
known magnitudes from the ''pg0231+051'' field of stars (Landolt  1992)  
with a color range $-0.3 \leq (B-V) \leq 1.5$ and $-0.1 \leq (B-I) \leq 3.0$ . 
Once the principal extinction coefficients in $B$, $V$, $R$ and $I$ were estimated,  
transformations of the instrumental magnitudes to a standard system were calculated 
according to the following equations:
\begin{eqnarray}
B-b = \alpha_{B} + \beta_{B}(b-v)_{0} \cr 
V-v = \alpha_{V} + \beta_{V}(b-v)_{0} \cr
R-r = \alpha_{R} + \beta_{R}(v-r)_{0} \cr
I-i = \alpha_{I} + \beta_{I}(v-r)_{0}, 
\end{eqnarray}
where $B$, $V$, $R$ and $I$ are the standard magnitudes, $b$, $v$, $r$ and $i$ 
are the instrumental (and airmass-corrected) magnitudes, and $\alpha$ and $\beta$ 
are the transformation coefficients for each filter.

A constant value associated to the sky background was subtracted using an interactive procedure
that allows the user to select regions on the frame free of galaxies and bright stars. 
Errors in determining the sky background, are, in fact, the dominant source of errors in 
the estimation of total magnitudes.

The most energetic cosmic-ray events were automatically masked using the COSMICRAYS task, and field stars were
removed using the IMEDIT task when necessary. Within the galaxy itself, care was taken to identify superposed stars.
A final step in the basic reduction involved registration of all available frames for each galaxy and in each filter to 
within $\pm 0.1$ pixel. This step was performed by measuring centroids for foreground stars on the images and then 
performing geometric transformations using GEOMAP and GEOTRAN tasks in IRAF.

\subsection{Errors}

Apparent magnitudes for each galaxy were estimated in three concentric circular 
apertures. This was achieved in $BVRI$ bands by using the PHOT routines in IRAF. 
Here we report the total apparent magnitudes while in the appendix  
we report apparent magnitudes at two other circular 
apertures, also in $BVRI$ bands. See Table A1.

An estimation of the errors in our photometry involves two parts: (1) The 
procedures to obtain instrumental magnitudes and (2) the uncertainty when such 
instrumental magnitudes are transformed to the standard system. 

For item (1), notice that the magnitudes produced at the output of the IRAF 
routines (PHOT) have a small error that is internal to those procedures. 
Since we have also   
applied extinction corrections to the instrumental magnitudes in this step, 
our estimation of the errors are mainly concerned with these corrections and 
the estimation of the airmass. After a least square fitting, the associated 
errors with the slope for each principal extinction coefficient are: 
$\delta(k_{B}) \sim 0.02$, $\delta(k_{V}) \sim 0.02$, $\delta(k_{R}) \sim 0.02$ 
and $\delta(k_{I}) \sim 0.015$. An additional error $\delta(airmass) \sim 0.005$ 
from the airmass routines in IRAF was also considered.

For item (2), the zero point and first order color terms are the most important 
to consider. After the transformation to the standard system by adopting our best--fit 
coefficients, the errors from the assumed relations for $\alpha$  were 
respectively 0.02, 0.04, 0.02 and 0.02 in $B$, $V$, $R$ and $I$, 
and 0.02, 0.03, 0.02 and 0.02 for $\beta$. To estimate the total 
error in each band, it is necessary to  propagate the errors, after considering the corresponding transformation 
equations.  An estimate of the sky contribution is necessary for quoting the total uncertainties. This was achieved 
by estimating the total magnitudes for all galaxies before and after sky subtraction. Typical values 
$\delta(B) \sim 0.08$, $\delta(V) \sim 0.08$, $\delta(R) \sim 0.9$ and $\delta(I) \sim 0.1$ are obtained. 
Total typical uncertainties are 0.1, 0.12, 0.12 and 0.15 in $B$, $V$, $R$ and $I$ bands, respectively.

The estimated total magnitudes in this work were compared with other external estimations when available 
in the literature. This has been done for: 1) The standard stars and 2) The isolated spiral galaxies.

\subsection{Standard Stars} 

For the standard stars, a comparison of our CCD magnitudes with those reported in Landolt (1992)
 for stars in the field of pg0231+051 and the Dipper Asterism M67 Star Cluster (Chevallier \& Ilovaisky 1991) 
are shown in Figure 1.


\begin{figure}
\plotone{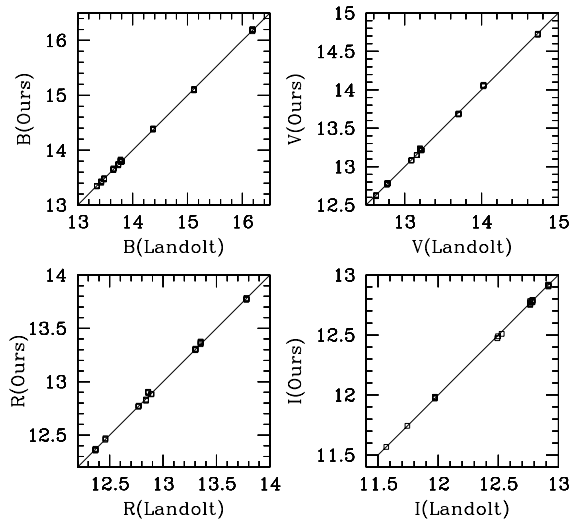}
\caption{Comparison between our estimated magnitudes
and those reported in Landolt (1992) for  standard stars in the field of pg0231+051 and the Dipper 
Asterism M67 Star Cluster (Chevallier \& Ilovaisky 1991). \label{fig1}}

\end{figure}


\begin{figure}
\plotone{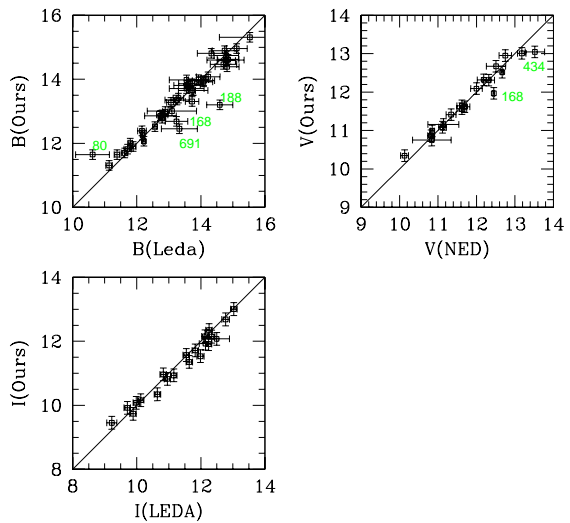}
\caption{Comparison between our total $B$, $V$, $R$ and $I$ magnitudes 
and the available photometry of similar aperture from the HyperLeda Database. Discrepant cases like CIG 80, 168, 
188, 434, and 691 are indicated. \label{fig2}}
\end{figure}

\begin{figure*}
\plotone{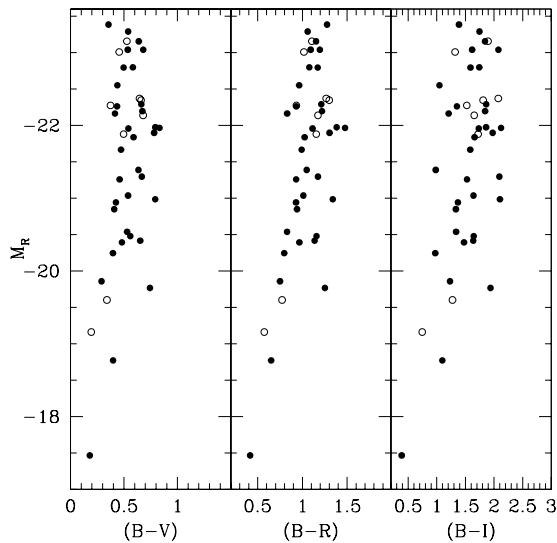}
\caption{Color--magnitude diagrams for the 44 isolated spirals after 
galactic and internal extinction corrections. Galaxies with 
inclination larger than $80^{\rm o}$ are showed with (open) 
circles.\label{fig3} }

\end{figure*}

Figure 1 shows no significant deviations between our CCD magnitudes and those reported for
the standard stars. A linear fit to this plot indicates a $\sigma \sim 0.005$ as the typical internal error for 
our magnitude estimations.

\subsection{Isolated Galaxies}

Figure 2 shows a comparison of the estimated magnitudes for the isolated galaxies in $B$, $V$ 
and $I$ bands versus the available total magnitudes in the Nasa Extragalactic 
Database (hereafter NED)\footnote{nedwww.ipac.caltech.edu} and aperture photometry in HyperLeda Databases. 
Discrepant cases such as CIG 80, 168, 188, 434, and 691 in the different bands are emphasized in the figure.
  

We find a reasonable agreement with the available values from the literature, except for a few 
discrepant cases shown in the Figure. HyperLeda reports detailed aperture photometry in the B-band for CIG 80. 
From a plot of the available data, the reported value should correspond to an aperture size  log (A) $\sim$ 2  while 
our magnitude corresponds to  log (A) = 1.58. The corresponding magnitude for an aperture similar to ours 
is 11.60 mag, in complete agreement with our estimation.

HyperLeda does not report any aperture photometry in the B-band for CIG 168. A detailed aperture photometry in 
the V and I (Cousins) bands is found instead. The V-band value in HyperLeda corresponds to 
log (A) = 1.44 while ours is log (A) = 1.50. An extrapolation of the available V and I 
magnitudes to our aperture size is consistent with our photometry. This makes us confident that our B-band value 
is well estimated and suggests that the B-band magnitude reported in HyperLeda should correspond to a 
smaller log (A) value. 
        
In the case of CIG 188, both HyperLeda and  NED 
databases report an homogenized magnitude from previously published data, 
assuming standard Johnson UBVRI filters. Notice however that NED magnitude is 14.10 $\pm 0.75$ with a large error bar. 
Given the absence of detailed aperture photometry and the diffuse nature of this galaxy (see corresponding images below), 
we suggest that the observed discrepancy is explained by the different aperture sizes. 

HyperLeda reports a couple of aperture data points for CIG 434 (corresponding to log (A) = 1.12 and 1.3, see 
Gallagher and Hunter 1986). By assuming a linear curve of growth, an extrapolation to log (A) = 1.5 (our aperture size) 
yields a magnitude value consistent with our data. NED reports an homogenized magnitude from previously published data, 
assuming standard Johnson UBVRI filters.

For CIG 691, HyperLeda reports a B--band magnitude of $13.33 \pm 0.562$ while NED reports a magnitude of
$12.60 \pm 0.5$ which is significantly closer to our reported value. Even more, the only aperture data point in 
HyperLeda suggests that the reported value corresponds to a smaller aperture size than ours.

Finally, the internal accuracy of our photometry was evaluated by comparing the total magnitudes 
derived from individual exposures. We find RMS differences between individual measurements of $\delta(B) \sim 0.06$, 
$\delta(V) \sim 0.06$, $\delta(R) \sim 0.05$ and $\delta(I) \sim 0.05$. Additional magnitudes at two other 
concentric circular apertures in $B$, $V$, $R$ and $I$ for all the isolated galaxies in this study are reported in 
the Appendix.

\section{Magnitudes and Colors}  
\label{Magnitudes}

The estimated apparent magnitudes and the colors of the galaxies in the sample are presented 
in Table 3. Entries are as follows: Column (1) gives the 
CIG number; Column (2) gives the logarithmic aperture size in 0.1 arcmin units, 
according to the HyperLeda convention, Columns (3) to (6) give the 
observed integrated apparent magnitudes in $B$, $V$, $R$ and $I$ bands. Finally, 
Columns (7) to (9) give the observed $(B-V)$, $(B-R)$ and $(B-I)$ color indices. 
Total typical uncertainties in our photometry are 0.10, 0.12, 0.11 
and 0.16 for $B$, $V$, $R$ and $I$ bands, respectively.

\placetable{tbl-3}

The $(B-V)$ corrected colors span the range of $0.2-1.3$ mag. 
This is comparable to that reported in other samples of non--interacting 
galaxies (e.g., de Jong 1996; Verheijen 1997). We emphasize important 
differences in blue Galactic absorption values between Burstein \& Heiles 
(1982) and Schlegel et al. (1998) for CIG 103 (1.485 vs 0.440 mag), 
CIG 138 (1.95 vs 0.65 mag), and CIG 144 (2.135 vs 0.510 mag).

The fraction of dust seems to be larger for bigger galaxies according to empirical 
(e.g., Giovanelli et al. 1995; Wang \& Heckman 1996; Tully et al. 1998) and theoretical (e.g., 
Shustov et al. 1997) arguments. Therefore, the internal extinction correction should depend 
not only on inclination but also on galaxy scale: $A_{\lambda}^i$[mag] = $\gamma_\lambda$log($a/b$), 
where $a/b$ is the major to minor axis ratio, and $\gamma_\lambda$ is a scale--dependent coefficient 
in the given passband $\lambda$.  From an empirical analysis, Tully et al. (1998) inferred the 
coefficients $\gamma_\lambda$ in the $BRIK$ bands as a function of the galaxy maximum circular velocity. 
From their data (given in Tully \& Pierce 2000) we have carried out linear correlations of these 
coefficients with the corresponding magnitudes {\it not corrected} for internal extinction:
\begin{eqnarray}
\gamma_B {\rm [mag]}= -6.30 -0.40 M_B, \ \ \ \  M_B < -16.7 \cr 
\gamma_R {\rm [mag]}= -4.20 -0.26 M_B, \ \ \ \  M_R < -17.7 \cr
\gamma_I {\rm [mag]}= -3.40 -0.20 M_I, \ \ \ \  M_I < -18.0 \cr
\gamma_K {\rm [mag]}= -0.85 -0.05 M_K, \ \ \ \  M_K < -19.7. 
\label{extin}
\end{eqnarray}
For values of the magnitudes larger than the limits given in eq. (\ref{extin}),
$\gamma_\lambda$ is assumed to be 0 (no extinction correction). For the band 
$V$, the line coefficients of $\gamma_V$ were obtained by a simple interpolation 
of those in the bands $B$, $R$, $I$, and $K$: $\gamma_V{\rm [mag]} = -4.67 -0.29 M_V$,
 $M_V < -17.5$. The $a/b$ ratios estimated at the
$B-$25mag/arcsec$^2$ isophote were taken from the HyperLeda Database.

Table 4 shows foreground and internal extinction corrected color indices and 
absolute magnitudes. Corrections are based on data generated from the dust Galaxy maps given in Schlegel 
et al. (1998) and available in NED database. Entries are as follows: Column (1) gives the identification 
CIG number; Columns (2) to (4) give the corrected $(B-V)$, $(B-R)$ and $(B-I)$ color indices. 
Finally, Columns (5) to (8) report the corrected absolute magnitudes in $B$, $V$, $R$ and 
$I$ bands. A Hubble constant value of 70 km s$^{-1}$ Mpc$^{-1}$ was adopted.

A more physical correction applied to the luminosities yields a B-band luminosity range 
($-18.1 \leq M_{B} \leq -22.25$) indicating no faint spirals in this sample, except for the 
case of CIG 434.

\placetable{tbl-4}

In Figure 3 we plot different Color--Magnitude Diagrams
[$M_{R}$ vs $(B-V), (B-R)$ and $(B-I)$ colors] for our sample of isolated galaxies. 
There is a mild correlation of colors with magnitude, although the sample is still small. 
In fact, a significant dependence of color on luminosity for normal isolated disk 
galaxies is not expected  (e.g., Avila--Reese \& Firmani 2000). 
The dependence that several papers reported in the past was mainly due to the dependence 
of the internal extinction on luminosity (or circular velocity, Tully et al. 1998), 
which we take into account accordingly here.


\section{Optical and Near-infrared (NIR) Morphology}
\label{Morphology} 

In order to discuss the optical morphology (that could be modified by the 
presence of bars, rings, etc. or external factors) and its relationship to 
the global photometrical properties, the images for each isolated galaxy are 
presented in the form of a mosaic in Figure 4 including, 
from upper-left to lower-right panel: 
1) a gray scale $R$-band image displayed at full intensity to look for faint 
external details; 2) an $R$ band sharp/filtered image to look for internal 
structure in the form of star forming regions and/or structure embedded into 
dusty regions; the filtered/enhancing techniques (Sofue 1993) allow the subtraction 
of the diffuse background in a convenient way for discussing the different morphological 
details; 3) a ($B-I$) color index map to visualize the spatial distribution of 
the SF (light--gray is for blue colors while dark--gray is for red colors); 
4) a composed (sharp/filtered) NIR $JHK$ image which is a combination of the 
archive $J, H$ and  $K-$band images from the Two-Micron Survey (Skrutskie et al. 
2006) to complement the structural and morphological analysis; and finally, 5) 
the ellipticity $\epsilon$ and Position Angle $PA$ radial profiles from the $I$ 
and composed $JHK$ images to provide evidence of the presence of bars and other 
structural details. (Figures 4.1 - 4.43 are available in 
the electronic edition of the Journal)
  
We identify a bar signature if the ellipticity radial profile $\epsilon$ rises to 
a maximum $\epsilon_{max}$ required to be above that of the outer disk, while 
the $PA$ radial profile shows a plateau (within $\pm 20^{o}$) along the bar 
(Wozniak et al. 1995).

All the images are oriented according to the standard (North-East) astronomical convention.   
The NIR images are approximately at the same scale as the optical images. For the sake of not crowding, the 
major diameter (arcmin) of the optical images is specified in the caption text for each galaxy.
In some cases, not all the foreground stars in each field have 
been removed.   


 \begin{figure*}
\vspace{12cm}
\includegraphics{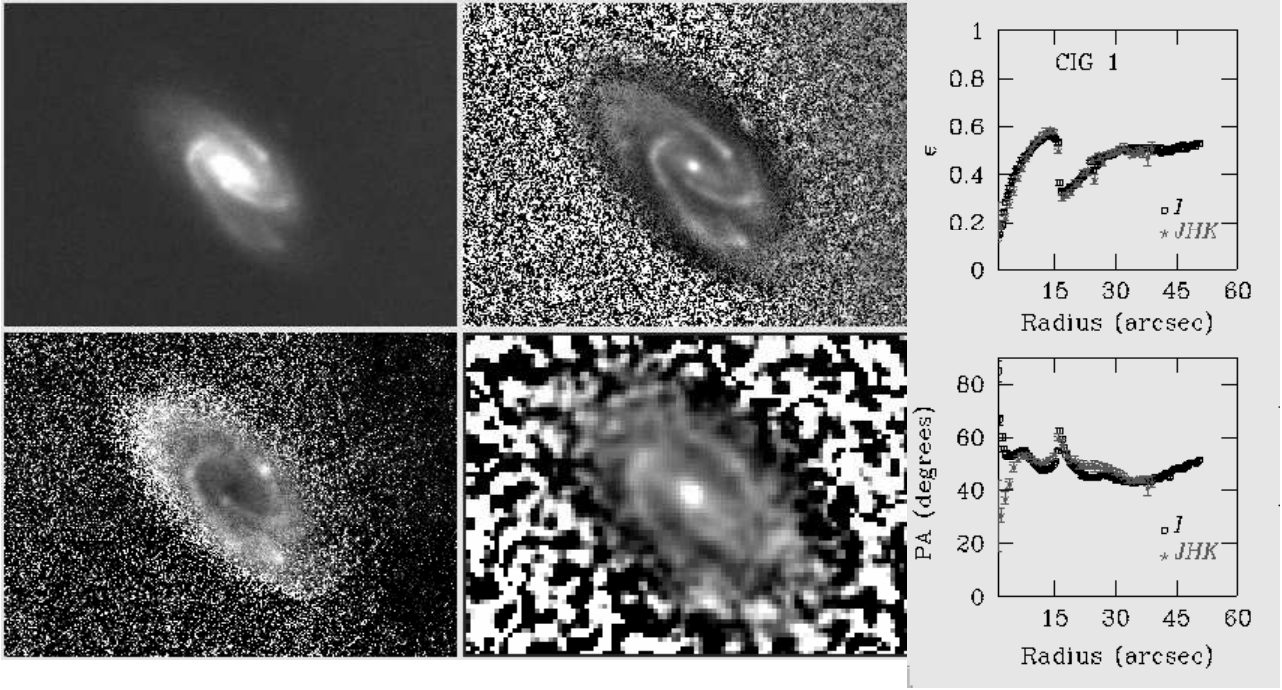}
\caption{CIG 1 Mosaic (example). Upper-left:a gray scale $R$-band image 
displayed at full intensity. Upper-right: An $R$ band 
sharp/filtered image. Lower-left: A ($B-I$) color index map. Lower-right:, 
 A composed (sharp/filtered) NIR $JHK$ image. Right-most panel: The photometric 
$\epsilon$ and $PA$ radial profiles from the $I$ and composed $JHK$ images. 
Images are oriented according to the astronomical convention. The 
major diameter of the galaxy in the optical images is 1.8 arcmin. \label{fig4}}
\end{figure*}


We use in addition the fact that the median value of the $(B-V)$ color declines systematically as the 
morphological type T increases along the morphological sequence. Median integrated total $(B-V)$ colors 
of galaxies according to morphological class are given by Roberts \& Haynes (1994). The UGC and the Local 
Supercluster (LSc) samples in Roberts \& Haynes (1994) are rather inhomogeneous in terms of environment, 
but the interacting objects were excluded from their analysis. The median colors of these samples will be 
used as a reference in the following discussion.

\subsection{Comments on Individual Objects} 
\medskip

{\bf CIG 1}. The galaxy was classified as SABbc (NED). The $R-$band and sharp/filtered images show a 
conspicuous fan-like structure at the end of the arms resembling a tidally disturbed galaxy. The optical 
images show three sharp-defined arms and a bar also visible in the composed JHK image. We classify 
this galaxy as SBbc. The total $(B-V)$ color is representative of Sc types. The photometric 
I and JHK band ($\epsilon$ and $PA$) profiles show evidence of a bar in this galaxy.

{\bf CIG 4}. The galaxy was classified as Sbc (NED). The gray-scale $R-$band image shows an inclined 
galaxy through a series of dust lanes. The optical R-band sharp/filtered and the $B-I$ color map images show   
two multiple and knotty arms and a strongly reddened central region. The composed JHK image shows two main arms 
and an elongated central barred-like region. We classify this galaxy as SABc. The total $(B-V)$ color is strongly 
reddened and not representative of Sc types. The photometric ($\epsilon$ and $PA$) profiles show evidence 
of a weak bar within the first 15 arcsec of this galaxy.

{\bf CIG 33}.  This galaxy was classified as SAB(rs)cd (NED). The gray-scale $R-$band image shows a faint outer 
arm in the east emphasizing its asymmetric appearance. Its internal s-shaped structure surrounded by a series of dust lanes 
and two prominent blobs in the north and south-east of the inner spiral arms are also apparent. The $B-I$ color index map shows 
a ring-like structure. The $JHK-$band image clearly shows an inner barred structure from which two arms emerge. We 
classify this galaxy as SB(rs)c. The total $(B-V)$ color is representative of Sc types. The photometric 
I and JHK band ($\epsilon$ and $PA$) profiles show evidence of a bar in this galaxy.

{\bf CIG 53}.  The galaxy was classified as a SB(rs)c (NED) . The gray-scale $R-$band and the sharp/filtered images show  
knotty features along ``multiple arms´´, a prominent bar encircled by a ring elongated in the direction perpendicular to the bar, 
and the presence of strong dusty structures. The  $B-I$ color map permits us to see a red central region and bluer arms. 
The JKH-band image confirms the barred and ringed structures and traces of the arms. We classify this galaxy as SB(r)bc. 
The $(B-V)$ color is representative of Sbc types. The photometric I and JHK band ($\epsilon$ and $PA$) profiles show evidence 
of a bar in this galaxy.

{\bf CIG 56}. The galaxy was classified as a SB(rs)b (NED). The gray-scale $R-$band image shows a barred galaxy with two
dominant arms and a third faint arm to the north-west. At the outskirts, the two arms become diffuse. The sharp/filtered 
image and $(B-I)$ color map show a blue ring elongated in direction of a red bar. The $JHK$ composed image shows a 
prominent bar and two dominant arms. We classify this galaxy as SB(r)b. The $(B-V)$ color is representative of Sb types.
The photometric I and JHK band ($\epsilon$ and $PA$) profiles show evidence of a bar in this galaxy.

{\bf CIG 68}.  The galaxy was classified as SAB(s)a (NED). The gray-scale $R-$band image shows a central elongated structure 
from which two arms emerge, while the sharp/filtered image shows traces of a bar. The  $(B-I)$ color map shows a 
bluer ring elongated in direction of the bar and an adjacent dust lane. The composed $JHK$ image shows a prominent 
bar and two arms. We classify this galaxy as SB(r)a. The $(B-V)$ color is representative of Sa types.
The photometric I and JHK band ($\epsilon$ and $PA$) profiles show evidence of a bar in this galaxy.

{\bf CIG 80}. The major axis of this galaxy is longer than our CCD frame. However we comment on the structure 
found within 4.3 arcmin. The galaxy was 
classified as a SA(s)b (NED). The gray-scale $R-$band and sharp/filtered images 
show a multi-armed and strongly asymmetric pattern enhanced by the presence of 
dust lanes, resembling a strongly perturbed system. The $(B-I)$ color map shows an inner ring and bluer arms. The 
composed $JHK$ image shows a barred-like structure from which two prominent arms seem 
to emerge and confirms the red nature of a ring oriented in the direction of the bar. 
We classify this galaxy as SB(r)b. The $(B-V)$ color is consistent with Sab types.
The photometric ($\epsilon$ and $PA$) profiles show weak evidence of a bar within the first 25 arcsec of 
this galaxy.

{\bf CIG 103}. The galaxy was classified as SAB(rs)c (NED) . This is another example of an apparently multi-armed 
galaxy seen through a series of strong dust lanes as shown in our $R-$band, sharp/filtered and $B-I$ color map images. 
The color map also shows an elongated red central region apparently encircled by a pseudo-ring alongside the 
direction of a bar. The composed $JHK$ image shows evidence of a weak bar from which two arms emerge. We classify this 
galaxy as SB(r)c. The $(B-V)$ color is consistent with Sc types. The photometric ($\epsilon$ and $PA$) profiles show 
weak evidence of a bar or an elongated structure within the first 15 arcsec of this inclined galaxy.

{\bf CIG 116}.  The galaxy was classified as RSB(s)a (NED). All our images show this galaxy with two inner well-defined 
arms making an s-shaped structure. These arms become diffuse as they wind out, giving the appearance of an 
external ring oriented almost perpendicular to the bar. The internal structure is seen through strong dust lanes. 
The composed $JHK$ image shows the presence of a bar and two inner dominant arms. We classify this galaxy as RSB(s)a. 
The $(B-V)$ color is consistent with Sa types. The photometric I and JHK band ($\epsilon$ and $PA$) profiles show evidence 
of a bar in this galaxy.

{\bf CIG 123}. This galaxy is classified as SB(rs)bc (NED). From the global distribution of optical light this galaxy 
appears asymmetric. Our images show an apparently multiple set of faint southern arms and a strong southern dust 
lane. The red central region, including the bar as well as the bluer ring and arms are emphasized in the $B-I$ color 
index map. The composed $JHK$ image clearly shows a strong bar, and a ring oriented in the direction of the bar. We classify 
this galaxy as SB(r)c. The $(B-V)$ color is consistent with Sbc types. The photometric I and JHK band ($\epsilon$ and $PA$) 
profiles show evidence of a bar in this galaxy.

{\bf CIG 138}. This galaxy was classified as SB(s)d (NED). This is an inclined galaxy seen 
through a series of dust lanes, causing the arms to appear multiple in nature. The $B-I$ color index map clearly emphasizes 
the reddened nature of the light distribution. In contrast, the composed $JHK$ image shows mainly two arms and an elongated 
central structure resembling a bar. We classify this galaxy as SBc.  The reddened $(B-V)$ color does not correspond 
to ScSd types. The photometric I and JHK band ($\epsilon$ and $PA$) profiles do not show evidence of a bar in this  
galaxy.

{\bf CIG 139}. The galaxy was classified as SB(s)m pec (NED). The $(B-I)$ color index map shows, however, a central 
reddened region and an outer bluer region similar to what is observed in spirals. There are no archive $JHK$ images for 
this galaxy. We preserve NED´s classification in this case. Notice that the obtained $(B-V)$ color is representative 
of Sm/Irr types. The photometric I-band ($\epsilon$ and $PA$) profiles do not show evidence of a bar.

{\bf CIG 144}. The galaxy was classified as Sb (NED). This is an edge-on galaxy showing a peanut-like-shape red 
bulge in the sharp/filtered image. However, the $B-I$ color index image and the 
composed $JHK$ image show a different structure. We classify this galaxy as SABb. The reddened 
$(B-V)$ color corresponds more to Sa types. The photometric I and JHK 
band ($\epsilon$ and $PA$) profiles are consistent with the presence of a bar in the first 30 arcsec of this inclined galaxy.

{\bf CIG 151}. The galaxy was classified as SAdm (NED). This is an inclined and apparently multi-armed spiral galaxy. The central
region appears elongated in the sharp/filtered image and red in the $B-I$ color map. The composed $JHK$ image suggests a bar 
structure and two adjacent arms. We classify this galaxy as SABc. The $(B-V)$ color corresponds to Sc types. The photometric I 
and JHK band ($\epsilon$ and $PA$) profiles show a weak evidence of a bar in this inclined galaxy.

{\bf CIG 154}. The galaxy was classified as SBcd (NED). This apparently disturbed galaxy shows an elongated and reddened 
central region. The winding of the inner arms suggests a ring oriented in the direction of the bar. 
The inner arms appear bifurcated at various places and faint outer arms are also perceived. The composed 
$JHK$ image shows a bar and a ring. We classify this galaxy as SB(r)cd. The $(B-V)$ color corresponds 
to ScdSd types. The photometric I and JHK band ($\epsilon$ and $PA$) profiles 
show evidence of a bar in the first 10 arcsec and of a ring at about 12 arcsec in this galaxy.

{\bf CIG 168}. The galaxy was classified as SAB(s)bc (NED). The galaxy appears moderately flocculent in the 
optical images with an elongated and red central region. The composed $JHK$ image also suggests an elongated 
central structure and only two prominent arms. The $(B-V)$ color corresponds to ScdSd types. We classify this 
galaxy as SAB(s)cd. The photometric I and JHK band ($\epsilon$ and $PA$) profiles do not show evidence of a bar.

{\bf CIG 175}. The galaxy was classified as SA(s)a pec (NED). The optical images show traces of diffuse and faint 
arm-like features and a prominent central region. The knotty appearance in the central part may be either intrinsic 
or caused by some coincident field stars not appearing in the composed $JHK$ image. The $(B-V)$ color corresponds 
to SmIm types. We classify this galaxy as Sa pec. The photometric I and JHK band ($\epsilon$ and $PA$) profiles are 
consistent with smooth precessing arm-like features.

{\bf CIG 180}.  The galaxy was classified as SA(rs)c (NED). The optical images show a multiple set of tightly 
wound arms and a reddened central region apparently encircled by an inner ring. In contrast, the composed 
$JHK$ image shows a prominent central region  and traces of the adjacent disk. We classify this galaxy as 
SA(r)b. This is an interesting case of a non-barred spiral with a symmetric inner ring that also shows 
hints of a circumnuclear ring. The $(B-V)$ color corresponds to Sab types. The photometric I and JHK 
band ($\epsilon$ and $PA$) profiles show weak evidence of a central ring within the first 10 arcsec and 
emphasize, in contrast with the optical image, a smooth nature of the central near-IR structure in this 
galaxy.

{\bf CIG 188}. The galaxy was classified as SAB(s)d (NED). This galaxy shows multiple arms in the optical images.
The sharp/filter image allows us to see a central bar structure that is red in the ($B-I$) color map. The low signal 
in the composed $JHK$ image does not permit us to appreciate the central bar but the photometric $I$ and $JHK$ 
band ($\epsilon$ and $PA$) profiles confirm the presence of a central bar. The total $(B-V)$ color is 
representative of SdSm types. We classify this galaxy as SB(s)d.

{\bf CIG 208}. The galaxy was classified as Sb (NED). This is a highly inclined galaxy. The sharp/filter and 
color index images show the central region resembling a long bar and the outer arms. The total $(B-V)$ color is 
representative of ScdSd types. The photometric $I$ and $JHK$ band ($\epsilon$ and $PA$) profiles also 
resemble a bar within the first 25 arcsec. We classify this galaxy as SABcd.

{\bf CIG 213}. The galaxy was classified as S0 (NED). We are using the $B$ band image to show a bar surrounded 
by an almost circular ring oriented in the direction of the position angle of the bar. The total $(B-V)$ color is 
representative of ES0 types. We classify this galaxy as RSB0. The photometric $B$ and $JHK$ band ($\epsilon$ 
and $PA$) profiles confirm a bar within the first 25 arcsec.

{\bf CIG 224}. The galaxy was classified as SB(rs)d (NED). The optical images show a galaxy of
 flocculent appearance with an outer pseudo-ring and evidence of a bar. The composed $JHK$ image is of low signal 
but still shows an elongated central region. The total $(B-V)$ color is representative of SdSm types. We classify 
this galaxy as RSBd. The photometric $I$ and $JHK$ band ($\epsilon$ and $PA$) profiles show a bar within 
the first 15 arcsec.

{\bf CIG 237}. The galaxy was classified as Sc (NED). This is an edge-on galaxy. The total $(B-V)$ color is 
representative of Scd types. Based on the observed prominence of the bulge region in the optical and composed 
$JHK$ images, we assume a Sc classification.

{\bf CIG 309}. The galaxy was classified as SA(r)ab (NED). The optical and $JHK$ images show an inner and outer 
ring. The optical images show an intermediate region of strongly wound and knotty arms?.  The total $(B-V)$ color 
is representative of S0Sa types. We classify this galaxy as RS(r)a. The smooth behavior of the photometric $I$ and 
$JHK$ band ($\epsilon$ and $PA$) is consistent with an early-type galaxy.

{\bf CIG 314}. The galaxy was classified as SAB(rs)c (NED). This is a multi-armed knotty spiral showing a red 
central region. The composed $JHK$ image barely shows part of two arms emanating from an elongated central 
region. The photometric $I$ and $JHK$ band ($\epsilon$ and $PA$) profiles show evidence of a weak bar. 
The total $(B-V)$ color is representative of ScSd types. We classify this galaxy as 
SAB(rs)c.

{\bf CIG 434}. The galaxy was classified as Im (NED). The optical images show a Magellanic-type irregular 
with an elongated and red region resembling a bar. There are no detected images available in the 2MASS 
survey. The total $(B-V)$ color is representative of Im types. We keep the Im classification. The photometric 
$I$ and $JHK$ band ($\epsilon$ and $PA$) profiles show evidence of a bar.

{\bf CIG 472}. The galaxy was classified as SAB(rs)c (NED). The optical images show a pattern of two arms 
that become bifurcated at the outer regions. In contrast, the composed $JHK$ image shows a single prominent 
northern arm extending to the south. The total $(B-V)$ color is representative of ScSd types. The photometric 
$I$ and $JHK$ band ($\epsilon$ and $PA$) profiles show no evidence of a bar. We classify this 
galaxy as SA(rs)c.

{\bf CIG 518}. The galaxy was classified as SA(s)c (NED). The optical images show a pattern of two arms 
that become bifurcated at the outer regions and a red elongated central region. In contrast, the composed $JHK$ 
image shows only two prominent arms that appear to enclose an elongated central region. The photometric 
$I$ and $JHK$ band ($\epsilon$ and $PA$) profiles show evidence of a bar. 
The total $(B-V)$ color is representative of SbSc types. We classify this galaxy as SB(s)bc.

{\bf CIG 528}. The galaxy was classified as SAbc (NED). The optical images show a spiral pattern with two
faint thin outer arms in the NE and SW resembling tidal tails. The composed $JHK$ image shows only part of 
two central arms. The total $(B-V)$ color is representative of ScSd types. The photometric 
$I$ and $JHK$ band ($\epsilon$ and $PA$) profiles do not show evidence of a bar. We classify this galaxy as 
SA(rs)cd.

{\bf CIG 549}. The galaxy was classified as SA(rs)c (NED). The optical images show a multiple set of arms while 
the composed $JHK$ image shows two arms at the center and an intriguing feature in the SE. The total $(B-V)$ color 
is representative of Sbc types. The photometric $I$ and $JHK$ band ($\epsilon$ and $PA$) profiles show 
evidence of a bar. We classify this galaxy as SB(rs)c.

{\bf CIG 604}. The galaxy was classified as (R)SB(s)a (NED). The optical images show a boxy bulge and two arms 
forming an outer ring. The sharp/filter and composed $JHK$ images show enhanced emission at the starting region 
of the arms in the NE and SW. The photometric $I$ and $JHK$ band ($\epsilon$ and $PA$) profiles show a large scale 
bar. The total $(B-V)$ color is representative of Sab types. We classify this galaxy as RSB(s)a pec.

{\bf CIG 605}. The galaxy was classified as SB(r)ab (NED). The optical images show multiple arms emanating from 
the end points of a bar. An inner ring oriented along the major axis of the bar is also appreciated. Two bright knots 
at the opposite sides of the bar are also seen in the composed $JHK$ image. The total $(B-V)$ color is representative 
of SabSb types. We classify this galaxy as SB(r)b.

{\bf CIG 691}. The galaxy was classified as SB(rs)d (NED). The optical images show a bar from which a diffuse set 
of arms emanates. At the composed $JHK$ image, only the bar can be seen. The total $(B-V)$ color is representative 
of ScScd types. We classify this galaxy as SB(rs)cd.

{\bf CIG 710}. The galaxy was classified as SA(s)cd (NED). The optical and composed $JHK$ images show an elongated 
central region resembling a bar. The photometric $I$ and $JHK$ band ($\epsilon$ and $PA$) profiles also show a weak 
evidence of a bar. The total $(B-V)$ color is representative of SdSm types. We classify this galaxy as SB(s)cd.

{\bf CIG 889}.  The galaxy was classified as Sa (NED). This is an edge-on galaxy showing a prominent peanut-shaped bulge from 
the optical to the NIR. This shape has been interpreted as representing a barred structure seen at high inclinations 
(c.f. Bureau \& Athanassoula 2005). We classify this galaxy as SBa. The $(B-V)$ color corresponds to Sa types. The photometric I 
and JHK band ($\epsilon$ and $PA$) profiles are consistent with the presence of a bar in the first 30 arcsec of this inclined 
galaxy.

{\bf CIG 906}.  The galaxy was classified as Sb (NED). This is an edge-on galaxy. The sharp/filtered image allow us to 
see a prominent dust band along the plane and the $B-I$ color index image shows this plane as highly reddened but with 
a bluer outer region. A bulge can be appreciated in the composed $JHK$ image. We preserve NED classification. 
The $(B-V)$ color corresponds to Sab types. The photometric I and JHK band ($\epsilon$ and $PA$) profiles may be 
consistent with the presence of a bar in the first 30 arcsec of this galaxy.

{\bf CIG 910}. The galaxy was classified as SBab (NED). This is another highly inclined galaxy that shows two main diffuse 
arms in the optical images. The sharp/filtered and $B-I$ color index images show evidence of a red barred structure and an outer 
bluer region. The composed NIR image confirms both the barred structure and the presence of two arms. We classify this 
galaxy as SBab. The $(B-V)$ color corresponds to Sa types. The photometric I and JHK band ($\epsilon$ and $PA$) profiles are 
consistent with the presence of a bar in the first 30 arcsec of this galaxy.

{\bf CIG 911}. The galaxy was classified as SBb (NED). The optical images show multiple arms emanating 
from a central ringed region that encloses a bar. The bar is red in color while the arms are bluer. The composed 
$JHK$ image permits us to see a barred structure enclosed by a ring oriented in the direction of the bar. We classify 
this galaxy as SB(r)b. The $(B-V)$ color corresponds to Sb types. The photometric I and JHK band ($\epsilon$ and $PA$) 
profiles show evidence of a bar in the first 15 arcsec of this galaxy.

{\bf CIG 935}. The galaxy was classified as SAB(rs)cd (NED). The optical images show a multi-arm pattern with an elongated
central region. In contrast, the composed $JHK$ image shows traces of a two-arm central pattern. The photometric $I$ and 
$JHK$ band ($\epsilon$ and $PA$) profiles show weak evidence of a bar. The $(B-V)$ color corresponds to SbcSc types. 
We classify this galaxy as SAB(rs)c.

{\bf CIG 976}. The galaxy was classified as Sab (NED). The optical images show an apparently disturbed spiral pattern 
while the composed $JHK$ image shows only a two-arm pattern. The photometric $I$ and  $JHK$ band ($\epsilon$ and $PA$) 
profiles do not show evidence of a bar. The $(B-V)$ color corresponds to SabSb types. We classify this galaxy as SA(s)ab.

{\bf CIG 983}. The galaxy was classified as SAB(rs)c (NED). The optical images show a multi-armed spiral pattern with
a slightly elongated central region. The sharp/filter image shows only a two-arm pattern and also a slightly elongated  
central region. The photometric $I$ and  $JHK$ band ($\epsilon$ and $PA$) profiles show weak evidence of a bar. The 
$(B-V)$ color corresponds to ScdSd types. We classify this galaxy as SB(rs)cd.

{\bf CIG 1004}. The galaxy was classified as SB(s)c (NED). The optical images show a bar and a multi-armed spiral
pattern. The composed $JHK$ image shows only a two-arm spiral pattern and a prominent large-scale bar. The photometric $I$ 
and  $JHK$ band ($\epsilon$ and $PA$) profiles also show evidence of a large bar. The $(B-V)$ color corresponds 
to SbcSc types. We classify this galaxy as SB(s)c.

{\bf CIG 1009}. The galaxy was classified as Sa (NED). The optical images show a set of tightly wound arms forming an 
inner ring. The composed $JHK$ image also shows a central arm pattern. The photometric $I$ and $JHK$ band ($\epsilon$ 
and $PA$) profiles do not show evidence of a bar. The $(B-V)$ color corresponds to SabSb types. We classify this galaxy 
as S(r)ab.

{\bf CIG 1023}. The galaxy was classified as SB(r)b (NED). The optical images show arms forming an outer ring. 
A bar is seen enclosed by an inner ring alongside the major axis of the bar. The arms appear to start at the 
end parts of the bar. The composed $JHK$ image shows prominence of the bar. The $(B-V)$ color corresponds to 
SabSb types. We classify this galaxy as RSB(r)b.

\subsection{Generalities of the sample}

Table 5 is a summary of the morphological results found 
in this work. Column (1) gives the original catalogue number, Column (2) 
gives the Hubble Type as reported in NED, Column (3) gives the Hubble Type 
as estimated in this work, Column (4) remarks the presence of Bars/Rings, 
Column (5) remarks the presence of multiple arms from the optical ($BVRI$) 
images, and Column (6) reports the Bar ellipticity (corrected for inclination).

\placetable{tbl-5}

NED database contains morphological information on subtypes for 
almost all these isolated galaxies, except for some of the most inclined ones (CIG 144,
208, 237, 889 and 906). 42$\%$ of the galaxy sample is earlier than Sbc and 52$\%$
is of Sbc type or later. The catalogue information concerning bars (confirmed 
and presumed) comprised 27 galaxies before this work, and we were able to add 
this information to 8 other galaxies.  This indicates that up to $79.5\%$ of the isolated 
galaxies in this subsample shows evidence of barred structure: for 63.5\% the
evidence is clear (SB galaxies) and for 16\% the bars are weak or suspected
(SAB galaxies). The bar fraction is roughly the same for early and late types. 
We have measured the $I-$band and $JHK$ isophotal ellipticities associated with a 
bar and calculated the maximum ellipticity, $\epsilon_{\rm max}$. This quantity 
(corrected by inclination) is related to a measure of the bar strength such as the gravitational 
bar torque (Laurikainen et al. 2002). 
Column 6 in Table 5 gives the values of $\epsilon_{\rm max}$ 
for our sample.  
Among barred galaxies, the average value of  $\epsilon_{\rm max}$ is $0.39\pm 0.1$. 
If we include the values of $\epsilon_{\rm max} = 0.39$, 7 early type galaxies 
and 6 late type galaxies have $\epsilon_{\rm max}\ge 0.4$, which is commonly considered as 
evidence of strong bar. Similarly, the catalogue information for rings in our sample 
comprised before 16 galaxies; in this work, it has been added to 8 other galaxies, accounting now to 
$55\%$ of the sample.

Finally, we emphasize the finding of clear morphological signatures of disturbance
at least in two galaxies, CIG1 and CIG80. The disturbance is revealed in the 
former case by a broad fan--like shape of the outer arms, while in the latter case 
it is revealed by a strong global asymmetric pattern of the multi--arms (see Figs. 4
and 10, and \S\S 4.1). 

\section{Physical Morphology}
\label{PhysMorphology}

Physical morphology has appeared as an alternative to classifying galaxies on the 
basis of physical properties rather than on visual features (see Morgan \& Osterbrock 
1969; Abraham et al. 1996; Conselice 1997; Bershady, Jangren \& Conselice 2000, 
among others). Conselice (2003, and more references therein) has provided a 
useful framework for classifying galaxies closely tied to underlying physical 
processes and properties. Conselice (2003; hereafter C2003) argues that the 
major ongoing and 
past formation modes of galaxies can be distinguished using three 
model--independent structural (photometric) parameters,
which allow for a robust classification system. These parameters are the 
concentration of stellar light (C), its asymmetric distribution (A), and a 
measure of its clumpiness (S).

Below we present below the $CAS$ parameters measured at various passbands for our
observed sample of isolated galaxies. The $CAS$ characterization of this set
of isolated galaxies is also helpful as a comparative sample for interpreting 
similar results of other surveys that sample galaxies in a wide range of 
environments (e.g., C2003; Hernandez--Toledo et al. 2005; 2006). 
Next, we briefly review each one of the $CAS$ parameters.

{\it Concentration of light $C$.-}
The concentration index $C$ is defined as the ratio of the 80\% to 20\% curve of 
growth radii ($r_{80}$, $r_{20}$), within 1.5 times the Petrosian inverted radius 
at r($\eta = 0.2$) ($r_P'$) normalized by a logarithm: 
$C = 5 \times log(r_{80\%}/r_{20\%})$ (see for more details C2003). 
The concentration is related to the galaxy light (or stellar mass) distributions.  

{\it Asymmetry $A$.-}
The asymmetry index is the number computed when a galaxy is rotated $180^{\circ}$ from its center and then subtracted 
from its pre-rotated image, and the summation of the intensities of the absolute value residuals of this subtraction is 
compared with the original galaxy flux (see for more details C2003). This parameter 
is also measured within $1.5\times r_P'$.  The $A$ index is sensitive to any feature that produces asymmetric 
light distributions. This includes galaxy 
interactions/mergers, large star-forming regions, and projection effects such as dust lanes (Conselice 1997; Conselice et al. 2000). 

{\it Clumpiness $S$.-}
Galaxies undergoing SF are very patchy and contain large amounts of light at high spatial frequency. To quantify this, the 
clumpiness index $S$ is defined as the ratio of the amount of light contained in high frequency structures to the total 
amount of light in the galaxy within $1.5\times r_P'$ (C2003). The $S$ parameter, 
because of its morphological nature, is sensitive to dust lanes and inclination (C2003).

 {\it Measurement of $CAS$ parameters.-}
The measurement of the $CAS$ parameters for the isolated spiral galaxies was carried out in several steps: 

(i) close field and overlapping stars were removed from each image; (ii) sky background was removed from the images; 
(iii) the center of each galaxy was considered as the barycenter of the light distribution and the starting point for 
measurements; (iv) the $CAS$ parameters for  all the spiral isolated galaxies were estimated directly, 
i.e. isolated galaxies are not influenced by light contamination from any other galaxy of similar size in the 
neighborhood (isolation criteria); (v) galaxies with high inclinations or axis ratios could introduce systematic 
biased trends in the values of the $CAS$ parameters (C2003). Isolated galaxies whose apparent axial ratios yield 
``inclinations'' larger than $80^\circ$ are represented as open circles on the corresponding plots.

\subsection{$CAS$ Results}

Since $CAS$ parameters are mostly reported in the $R$ band by other authors,
in order to compare, we provide here the calculated $CAS$ parameters and their errors, 
also in the $R$ band (Table 6). The $CAS$ values in the other bands for our observed 
isolated spirals will be provided by the authors upon request.

\placetable{tbl-6}

By sorting the sample in early-- and late--type spirals (SaSb and SbcSm, 
respectively), the 
corresponding average and standard deviation values of the $CAS$ parameters are: 
$<C(R)> (SaSb) = 4.00 \pm 0.50$,  $<A(R)> (SaSb) = 0.08 \pm 0.05$, 
$<S(R)> (SaSb) = 0.20 \pm 0.12$, and $<C(R)> (SbcSm) = 3.10 \pm 0.40$,  
$<A(R)> (SbcSm) = 0.19 \pm 0.10$, $<S(R)> (SbcSm) = 0.36 \pm 0.20$. Our mean 
values are consistent with those reported in C2003 for the Frei et al. (1996) 
sample of non--interacting galaxies, except in the case of $<S(R)>$ (SbcSm). 
Notice however that irregulars were included in in SbcSm class while in 
C2003 these galaxies are separated.    

\begin{figure}
\plotone{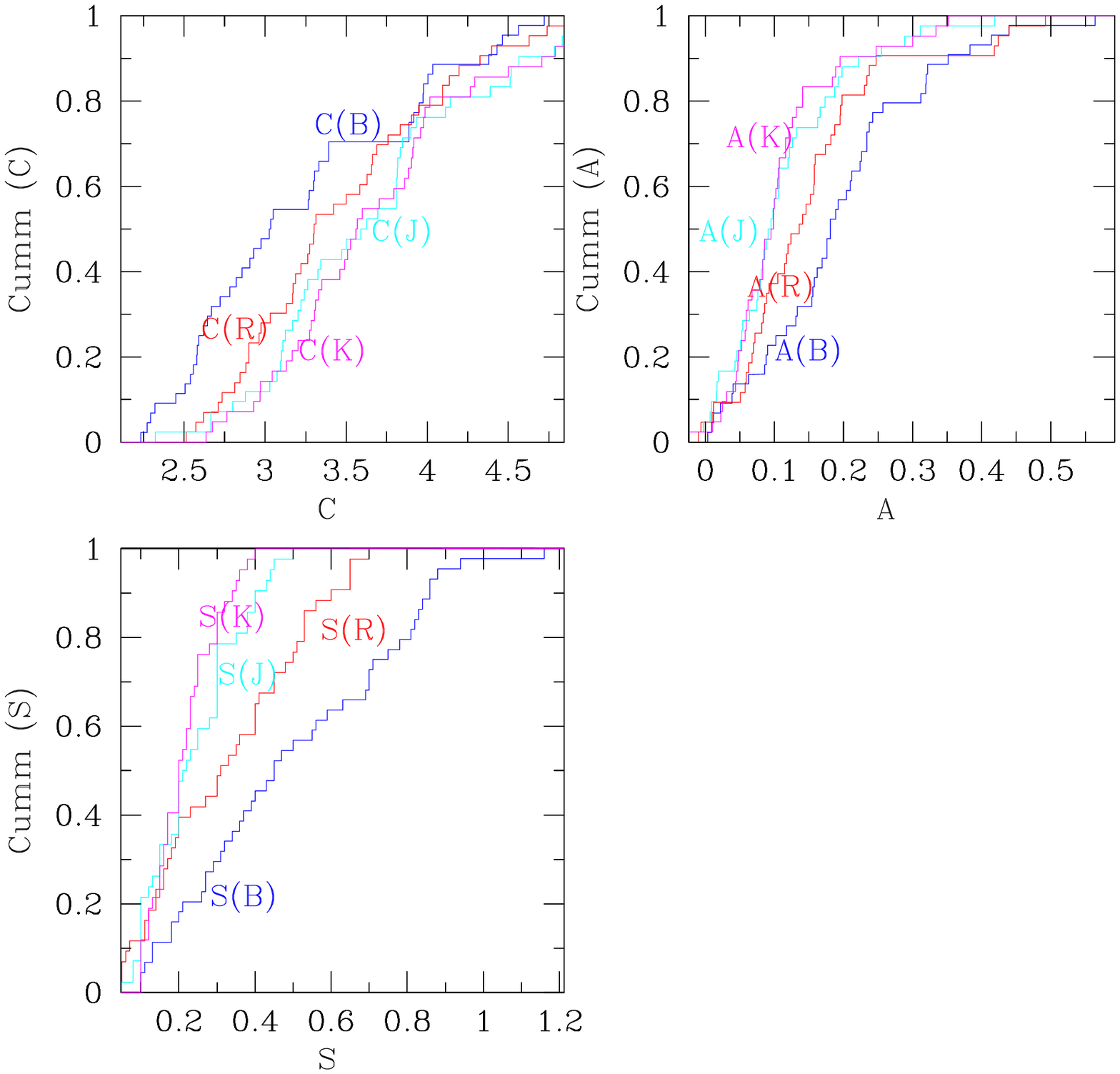}
\caption{Cumulative distribution function of $CAS$ parameters in the 
bands $B$, $R$, $J$ and $K$ bands. }
\label{cumdistr}
\end{figure}

An interesting question is how the $CAS$ parameters do change with
wavelength.  Figure \ref{cumdistr} shows the cumulative distribution function 
of the $CAS$ parameters at $B$, $R$, $J$ and $K$ bands. This comparison let 
us see visually that there are significant systematic changes in the 
$CAS$ parameters with wavelength. The concentration $C$ becomes higher 
from bluer to redder bands, specially for those galaxies with low and 
intermediate values of $C$. In the case of both the asymmetry $A$ and 
clumpiness $S$ parameters, their values strongly decrease from bluer to 
redder bands, more as larger are these parameters.  

\begin{figure}
\plotone{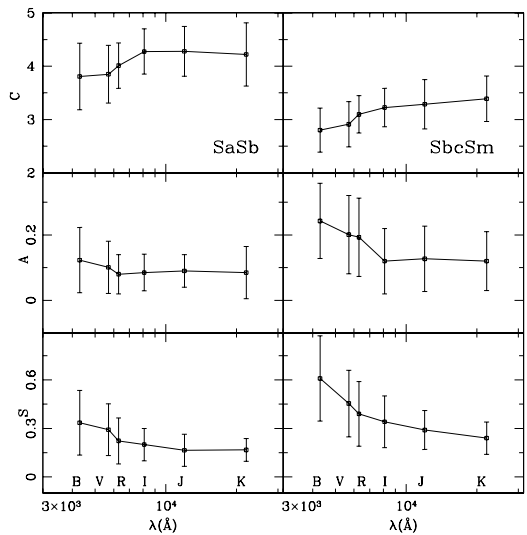}
\caption{Average and standard deviation values of the $CAS$ parameters
as a function of the central wavelength $\lambda (\AA)$. Left and right columns are
for the subsamples of early (SaSbc) and late (ScSm) isolated galaxies.}
\label{cas_band}
\end{figure}

In Figure \ref{cas_band} we plot the average and standard deviation values
of the $CAS$ parameters vs wavelength (color band) for our sample sorted
in early-- and late--type spirals (SaSb --left panel-- and SbcSm --right
panel, respectively). The $CAS$ parameters of later types show on
average more dependence (and scatter) with wavelength than the early
types. Among the $CAS$ parameters, the clumpiness is the most
sensitive to wavelength.

 
\section{Discussion} \label{S4}

\subsection{Morphology, Bars, and Rings in Isolated Galaxies}

The optical and IR emission in galaxies are dominated by different populations 
of stars and are subject to dust absorption at different levels. The structures that 
are dominated by older stellar populations are more prominent and less affected 
by extinction in the IR than in the optical. In the IR one observes a 
higher bar fraction than in the optical, as well as cases 
where the bulge appears more prominent, the spiral arms less flocculent and rings 
less prominent (Eskridge et al. 2000). 
The latter points towards an earlier--type classification from the IR images 
than from optical ones. The same trends mentioned above are observed for our 
sample of isolated galaxies. It should be remarked that, in our attempt to 
describe the morphology of the observed galaxies, we also have made use of the 
$JHK$ images from the the Two Micron-Survey. 
 
The re--classification presented here (see Table 5) preserves the optically 
observed morphology but takes into account the NIR bar morphology. 
In general, previous results concerning the differences in morphology
as passing from optical to NIR bands (Eskridge et al. 2000) agree with 
the ones seen in our subsample of isolated galaxies, though the fraction 
of isolated galaxies for which these differences become significant is 
actually small in our case ($\sim 15\%$).

Concerning bars in our isolated galaxies, they come into a variety of sizes, 
shapes and color distributions; from apparently strong, to small ones confined 
to the central parts of galaxies and up to the oval-shaped bulges, suggesting 
a range of strengths, lengths and mass distributions. We have shown that the 
fraction of galaxies in our sample with clear
evidence of optical/IR bars (SB galaxies) is 63\%, while 
16\% more show some evidence of weak bars (SAB galaxies). These
fractions are in agreement with estimates from larger samples
of galaxies. For example, Eskridge et al. (2000) determined the fraction 
of strongly barred galaxies in the $H-$band for a sample of 186 spirals 
{\it from different environments} to be 56\%, while another 16\% is weakly 
barred. For the same sample, the fraction of barred galaxies reported
in the optical is almost a factor of two smaller than in the NIR. 
We do not find such a strong difference with the passband in our sample of 
isolated galaxies. 
We also reported the presence of inner (r) and outer (R) rings when possible, 
but a detailed ring morphology (Buta 1986; 1995) was not attempted. The
fraction of galaxies with rings in our sample is high, 55\%.

Notice that the observed fraction of bars and rings in the present paper 
can hardly be a bias of our observing procedure since we simply selected 
objects according to their availability in the sky.

The high fractions of bars for the isolated 
galaxies found here, similar to the fractions observed in other environments, 
could be suggesting that interactions and the global effects of the 
group/cluster environment are not crucial for the formation/destruction 
of bars. How do bars form in isolated environments?  It is known that the 
presence and evolution of bars in a Hubble time depends on the host 
galaxy structure, the dark matter halo structure, the disk-to-halo ratio, 
as well as on the environment (e.g., Athanassoula 2003;  
Berentzen, Shlosman \& Jogee 2006; Col\'{\i}n et al. 2006). 
High--resolution N-body simulations of isolated disks embedded in CDM halos 
show that extended strong bars form almost always, but they slow down as a 
result of angular momentum transport to the disk and halo (Debattista \& 
Sellwood 2000; Athanassoula \& Misiriotis 2002; Valenzuela \& Klypin 2003);  
eventually, the bars may dissolve forming a pseudobulge (e.g., Avila-Reese 
et al. 2005; Berentzen et al. 2006). 

The isolated environment may be ensuring the presence 
of dynamically cold disks that can form a variety of stellar bars.  
The observed bar fraction could also be a consequence of long--lived bars, 
or alternatively, of bars that recurrently form, self--destroy, and resurrect
due to gas accretion (Bournaud \& Combes 2002). Constant gas accretion
is a condition more viable in isolated environments than in group/clusters.

Our data reveal no difference in the relative bar fraction of early-- (SaSb) 
and late-- (SbcSm) type galaxies. If any, among the barred galaxies, the 
late--type subsample contains a larger fraction of weakly barred (SAB) galaxies 
than the early--type subsample.  Eskridge et al. (2000) also found that the fraction
of barred galaxies almost does not depend on morphological type. It seems
that rather than the fractions, the properties of the bars (length, 
strength, surface brightness profile, etc.) are those that change as a function 
of the morphology and/or environment (e.g., Erwin 2005 and references therein). 
For our sample, we have estimated the bar deprojected 
maximum ellipticity, $\epsilon_{\rm max}$. We do not find significant differences
in $\epsilon_{\rm max}$ as a function of morphological type
as well as a function of the $CAS$ parameters. Further analysis is
necessary to infer the disk and bar properties and compare them with model
predictions.

Another relevant topic is that of ring formation in isolated environments. 
Rings of SF are a common phenomenon in disk galaxies. Most rings form by gas 
accumulation at resonances, usually under the continuous action of gravity torques 
from a bar pattern, but sometimes in response to a mild tidal interaction with a 
nearby companion (Buta \& Combes 1996; Buta 1999). In either case a resonance is 
a very special place in a galaxy where SF can be enhanced and may proceed either 
as a starburst or continuously over a period of time. 
Most of the observed rings in our galaxies are of the type encircling the 
end of the bars and 
elongated along the bar's position angle. Singular cases of outer rings, 
inner rings and a probable circumnuclear ring were detected. Contrary to bars, most of the 
observed rings in these isolated galaxies show bluer color distributions 
suggesting that their stellar populations are more similar to those in their 
hosting disks than those in the bars.  However, some of our composed $JHK$ images 
also show the prevalence of the rings in the NIR. The existence of this 
old population rings underlying star forming rings suggest a strong coupling 
between the stellar and gaseous components in the resonance regions.

Finally, it should be mentioned that in any of the 44 isolated galaxies studied here,
we did not find strong signatures of interactions or perturbations.
However, in two cases (CIG1 and CIG80) moderate morphological
distortions can be seen, which could evidence some level of dynamical disturbance, though, 
these distortions could hardly be produced by strong interactions, as is
the case of the isolated disturbed galaxies reported in Karachentsev et al. (2006,
see Introduction). Satellite accretion could explain the distortions seen
in CIG1 and CIG80. On the other hand, as mentioned above, bars and rings are 
axi--symmetric structures that can
be explained as a product of the internal secular evolution of disks. Interactions
and perturbations may induce and amplify bar/ring formation but can also 
contribute to their fast dissolution. A larger sample is necessary in order
to explore whether or not a fraction of isolated galaxies shows evident
signatures of interaction, a question of relevance as mentioned in the Introduction.

\subsection{Physical Morphology through the $CAS$ parameters}

Light concentration, asymmetry and clumpiness $CAS$ parameters have been used 
in alternative galaxy classification schemes (see for references \S 5).
The $CAS$ parameters allow also the possibility to classify galaxies
according to their interaction state (C2003; Hern\'andez--Toledo
et al. 2005,2006), which is useful in high--redshift studies. In this sense, 
it is important to have a well studied 'comparative' sample of local isolated 
galaxies. In spite of the small number
of galaxies in our current sample, we will introduce below an indicative
discussion of the measured $CAS$ parameters in different color bands and of 
their trends with other galaxy properties.

Figure \ref{casplanes} shows the loci of the isolated SaSb and SbcSm 
galaxies in the projected planes of the $R-$band $CAS$ space. Only the 
averages and their standard deviations are shown (crosses and continuous
error bars). The boxes indicate the amplitude of variation of the
$CAS$ parameters (lower and upper limits) from $B, 
V, R, I$, $J$ to $K$ bands. For comparison, the $R-$band averages and standard 
deviations of galaxies in interacting S+S pairs (Hern\'andez--Toledo et al. 2005), 
and starburst and Ultra Luminous Infrared (ULIR) galaxies (C2003) are also plotted. 
Visually, the major difference between isolated spirals and interacting, 
starburst and ULIR spirals takes place in the $A-S$ plane.  
 


\subsubsection{Concentration}

The quantitative measure of $C$ in our isolated spirals span the range 
$3.0 \leq C(R) \leq 4.5$, in agreement with other works. The average and standard 
deviation values are $<C(R)> = 3.5 \pm 0.59$. 
It is known that for lenticular and elliptical galaxies, the concentrations 
are typically larger than for spirals (e.g., C2003;
Hern\'andez--Toledo et al. 2006).   We have also found that $C$ systematically 
increases with the passband for almost each one of the isolated galaxies
(Fig. \ref{cumdistr}). This is in agreement with previous finds that the scalelength 
of spirals is smaller in NIR bands than in the optical ones (e.g., de Jong 1996), 
which is probably pointing out probably to an inside--out galaxy formation scenario:
bluer colors trace younger stellar populations, and if the disk is more extended
in the optical than in the NIR, then the outer disk might be younger
(more recently assembled) than inner regions. Metallicity gradients would 
emulate such an effect, but it seems that this is not the case (de Jong 1996).

According to our current understanding of galaxy formation, disks form 
generically inside--out within growing CDM halos (see for a recent review 
Avila--Reese 2006). Their concentrations (or surface brightnesses) depend 
mainly on the spin parameter of the halo. CDM halos span a wide lognormal 
distribution of the
spin parameter, hence, one expects also a wide range of concentrations 
for the disks. Probably, the observed distribution of concentrations
for isolated disk galaxies is not as wide as we would expect from theory. 
It should be also taken into account that internal secular processes after 
disk formation rearrange the mass (light) distribution, and that the presence
of a large bulge in early--type spirals tends to increase their $C$ parameter
with respect to galaxies with smaller bulges.  A fair comparison
of concentrations observed in isolated spirals with model predictions
is certainly a promising avenue of research. In the cases of
ellipticals (and probably the bulges of early--type galaxies), theory 
suggests that they are more concentrated due to 
the violent and dissipative processes that are at the basis of their
formation: major mergers of gaseous disks. 

Concentration of light has been shown to correlate with properties 
of galaxies like Hubble type, color, and surface brightness 
(Okamura et al. 1984; C2003; see below). 
More recent studies have also shown that concentration of light correlates 
with internal scaling properties such as velocity dispersion, size, 
luminosity, and black hole mass (Graham et al. 2001).

\subsubsection{Asymmetry}
 
Concerning asymmetry, the method used here gives a simple quantitative 
measure of how a galaxy deviates from axi--symmetry. The asymmetry parameter 
$A$ has been shown to be sensitive mainly to galaxy interactions/mergers, 
but also is influenced by SF clumps, dust lanes, and projection
effects. The quantitative measure of $A$ in the present sample of isolated galaxies 
spans roughly the range $0.0 \leq A(R) \leq 0.23$, the average and
standard deviation being $<A(R)> = 0.15\pm 0.10$. The later types are
slightly more asymmetric on average than the earlier ones. The asymmetries
reported here are definitively lower than the typical ones of interacting 
disk galaxies (see Fig. \ref{casplanes}). The $A$ parameter decreases 
significantly as the passband is redder (Fig. \ref{cumdistr}).
For interacting spirals, the same trend was observed, but in a much
less extent (Hern\'andez--Toledo et al. 2005). This suggests that 
while the (high) asymmetry in interacting spirals is mainly of
global/external origin --hence is more or less the same in different bands--, 
in the case of isolated spirals the (low) asymmetry is in part related 
to SF effects; this is why $A$ is so sensitive to the passband 
in which it is measured. The question of whether a disk of a spiral 
galaxy is intrinsically asymmetric or not is of great interest. Some studies 
have shown that important deviations from axisymmetry exist in the optical 
and other wavelengths (Rix \& Zaritsky 1995; Richter \& Sancisi 1994; 
C2003). However, systematic attempts to quantify asymmetry and other 
measures like the $CAS$ parameters in several wavelengths for well--selected 
local samples of galaxies are either rare or missing in the literature.

\subsubsection{Clumpiness}

\begin{figure*}
\vspace{14cm}
\includegraphics{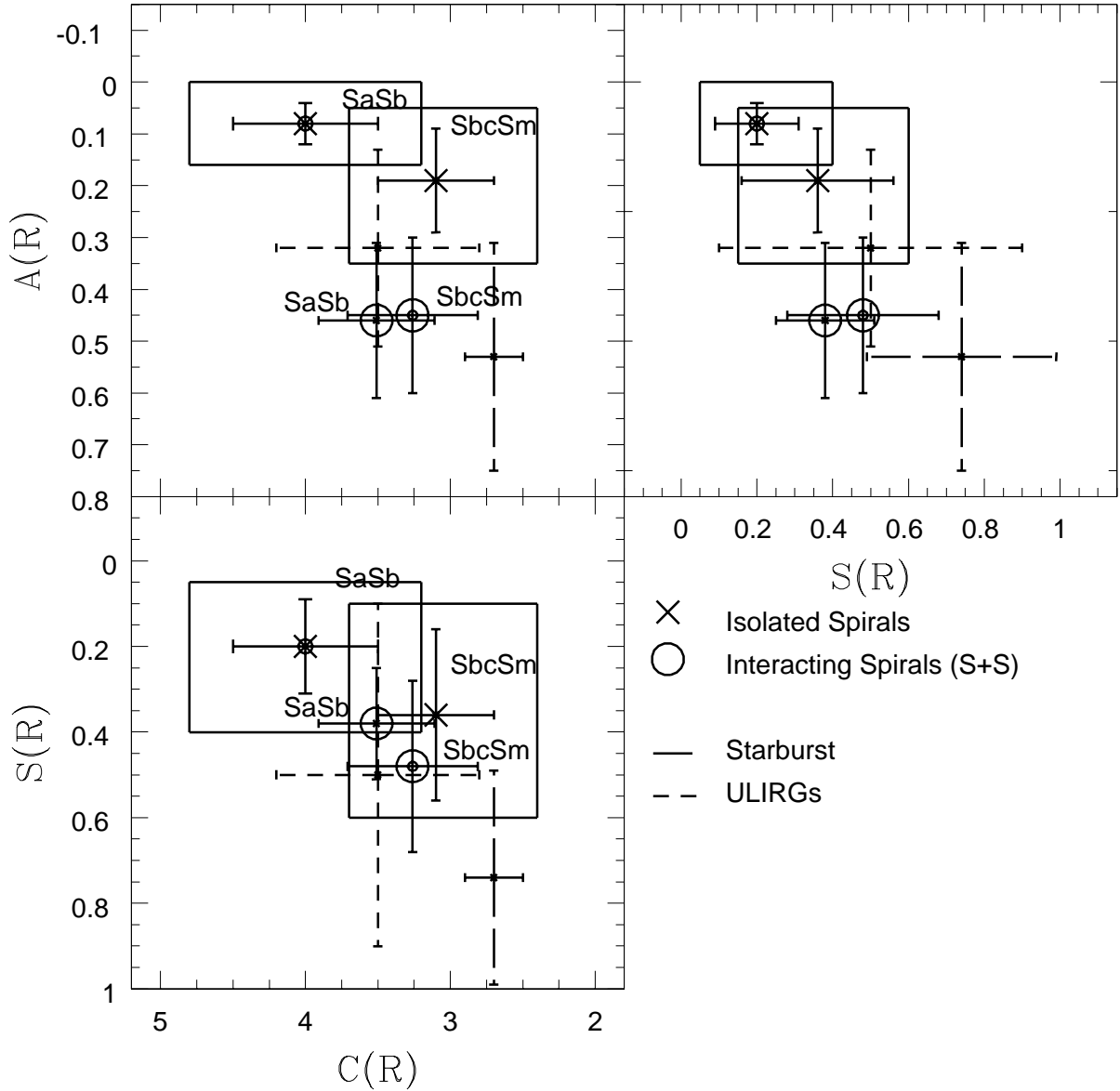}
\caption{Loci of the mean $R-$band $CAS$ values and their 1$\sigma$ dispersion for 
our isolated SaSb and SbcSm galaxies in the $CAS$ planes (crosses with error
bars). The large boxes illustrate the amplitude of variation of the $CAS$ values 
from all the bands ($B$, $V$, $R$, $I$ to $J$ and $K$). 
The corresponding $R-$band values for the interacting
SaSb and SbcSm galaxies are shown with circles and solid error bars.
(Hernandez--Toledo et al. 2005).
Short--dashed and long--dashed error bars are for ULIR and starburst
galaxies, respectively (C2003).  }
\label{casplanes}
\end{figure*}

Galaxies undergoing SF are patchy, specially in the bluer 
bands, and an important fraction of light must be in high spatial frequency 
structures. This is quantified through the clumpiness $S$ parameter. 
For our sample of isolated galaxies, $S(R)$ ranges roughly from 
$0.0$ to $0.6$, being the average and standard deviation values
$<S(R)> = 0.31\pm 0.15$. The $S$ parameter is on average larger and more 
scattered in later types than in the earlier ones as is seen in Fig. \ref{casplanes}
(see also below). It is well known indeed that late type 
galaxies present more current SF activity than the early type ones. 
Although, the parameter $S$ in our isolated galaxies is typically smaller 
than in interacting spirals, the differences 
are actually small, ant not so significant as in the case of the asymmetry 
parameter (Fig. \ref{casplanes}). It is also interesting to note 
that the increase of the $S$ value as the passband is bluer in our 
isolated galaxies (Fig. \ref{cumdistr}) is much more significant 
than in the case of interacting galaxies (Hern\'andez--Toledo et al. 2005).

\subsubsection{Correlations}

\begin{figure}
\plotone{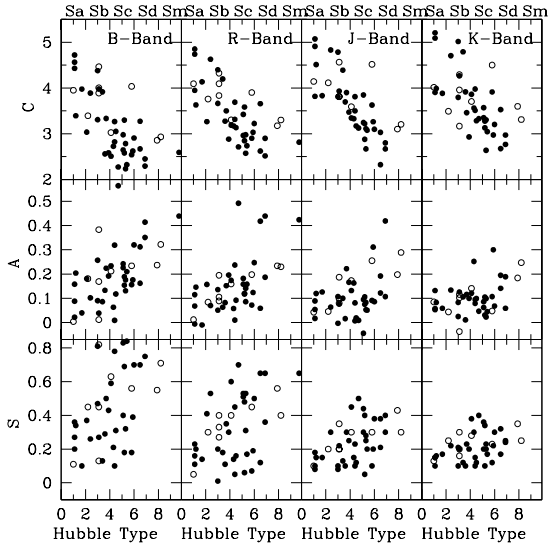}
\caption{$CAS$ parameters in the $B$, $R$, $J$ and $K$
bands versus the Hubble type. Galaxies with inclination 
larger than $80^{\rm o}$ are showed with (open) circles. }
\label{cas_T}
\end{figure}

\begin{figure}
\plotone{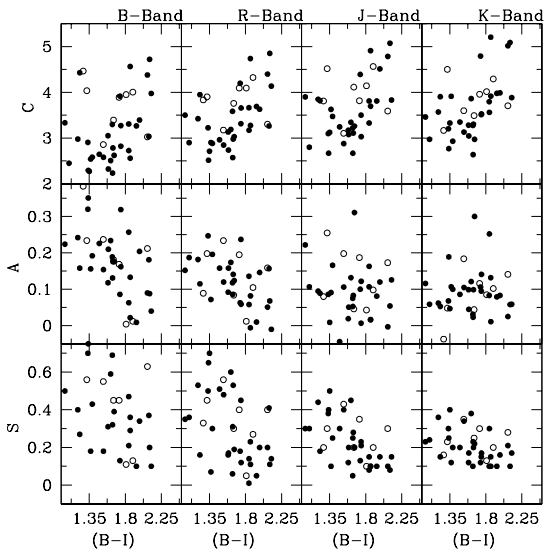}
\caption{$CAS$ parameters in the $B$, $R$, $J$ and $K$ 
bands versus the corrected total $(B-I)$ color. Galaxies with 
inclination larger than $80^{\rm o}$ are showed with (open) 
circles.}
\label{cas_color}
\end{figure}

We next explore how the $CAS$
parameters of isolated galaxies correlate with other properties and whether 
these correlations are sensitive to the passband or not. Figures \ref{cas_T} 
and \ref{cas_color} show the $B$, $R$, $J$ and $K$ band $CAS$ 
parameters vs morphological type $T$ and corrected total $(B-I)$ color.
Nearly edge--on galaxies (inclination $\ge 80^{\rm o}$) are plotted with
(open) circles. Given that $A$ and $S$ are particularly sensitive to 
projection effects, it is important to visualize these galaxies since 
they may be masking any trend.

After a visual inspection of Figs. \ref{cas_T} and \ref{cas_color},
the general conclusion is that any potential trend of the $CAS$ parameters
with $T$ and total $(B-I)$ color tends typically to be more robust in the redder 
bands. This emphasizes 
the merits of IR parameters, which are less contaminated from (transient) 
SF effects and better represent the basic structure of galaxies. 
We should note that the scatters in these trends, even for the $J-$ band, 
are large. The images from the Two--Micron Survey are unfortunately
of low quality, specially in the $K-$band. Besides, for several of our
galaxies, there are not images in this Survey. Therefore,
the $J$ and $K$ band data discussed here should be taken only
as indicative ones. 

Larger samples are needed in order to quantify better the showed
dependences in Figs. \ref{cas_T} and \ref{cas_color}, and infer from them clues to the physics 
of disk galaxies. Notwithstanding this, we present below a brief discussions 
on the observed trends.

According to Figs. \ref{cas_T} and  \ref{cas_color}, while the 
concentration tends to be higher for earlier--type and redder galaxies, 
the asymmetry and clumpiness tend to become smaller.
The morphological type is led mainly by the bulge--to--disk ratio. 
The global color is also affected by this ratio.
Therefore, it is expected that earlier types be more concentrated and
redder. However, the $C$ parameter and the global color are not 
too sensitive to the bulge--to--disk ratio for galaxies with intermediate-to-small 
values of this ratio (say Sb types and latter); therefore, in 
these cases, the measured $C$ and global color reflect mostly the pure 
disk concentration and color. Thus, that $C$ depends on $T$ for late types, 
implies mainly a connection between the spiral arm properties and the disk 
concentration. The dependence of $C$ on color would imply mainly that less 
concentrated disks have a more constant SF history, probably 
because their gas surface densities are low.

Concerning asymmetry, the observed dependence on $T$ indicates that most 
of the asymmetry of our isolated spirals is associated with the natural 
flocculency in later-type galaxies as well with SF, which is
more active for later types (as is evidenced also by the trend of higher
$S$ values as the types are later and the colors bluer, see above). In this 
interpretation, the effect of large--scale perturbations (c.f. interactions) 
is neglected. The $A$ parameter could be used as a first--approximation 
indicator of interaction signatures in isolated spirals. An automatic
analysis of images in large samples of galaxies provides easily the 
$A$ parameter. An even more reliable test for interactions would be
to produce the loci of the studied isolated spirals in the $A-S$ diagram,
where maximum differences between isolated and interacting galaxies
are revealed as was seen above (Fig \ref{cas_T}; see also
Hern\'andez--Toledo et al. 2005). The adding of
color information is also valuable as evidenced by Fig \ref{cas_color}.


\section{Summary and Conclusions} \label{S5}

We present results of our $BVRI$ CCD photometry for a set of 44 
isolated galaxies selected from the CIG catalogue (Karachentseva 1973). 
We have shown that our derived parameters are generally in good agreement with 
the aperture photometry reported in the HyperLeda database and other individual 
photometric works. 
In addition, we present multiaperture photometry (Appendix) in order to 
facilitate further comparisons and contribute to the 
existing databases of aperture photometry (e.g., HyperLeda). 

In a further step, we have analyzed the morphology of each one of the 
galaxies based on our mosaic $R-$band, sharp/filtered 
$R$ band images, 2D $(B-I)$ color maps, composed 
NIR $JHK$ images from the Two--Micron Survey archives and photometric $\epsilon$ 
and $PA$ radial profiles. A morphological
re--classification of the galaxies has been presented with emphasis on
structural features as bars and rings and global disturbances.

The sample morphological types range from Sa to Sm, half of the 
galaxies being earlier than Sbc (SaSb), and the other half being Sbc or later 
(SbcSm). After our reclassification, we have found that $\sim 63\%$ of 
the galaxies is clearly barred (SB), while a $\sim 17\%$ more shows some 
evidence of a weak bar (SAB). There is not any significant difference
of the bar fraction with the morphological type. The average and
standard deviation values of the $I-$band deprojected
maximum ellipticity of the bars, $\epsilon_{\rm max}$, is $0.39\pm 0.1$.
There is not any trend of $\epsilon_{\rm max}$ with the morphological
type and the $CAS$ parameters. We have also found that 55\% of the isolated
galaxies in our sample shows ring structures.

Finally, we have calculated the $BVRI$, $J$ and $K-$ band concentration, asymmetry, and 
clumpiness ($CAS$) parameters for the sample. The $CAS$ averages and standard 
deviations in the $R-$band for the SaSb and SbcSm subsamples are: 
$<C(R)> (SaSb) = 4.00 \pm 0.50$, $<A(R)> (SaSb) = 0.08 \pm 0.05$, 
$<S(R)> (SaSb) = 0.20 \pm 0.10$ and $<C(R)> (SbcSm) = 3.10 \pm 0.50$,  
$<A(R)> (SbcSm) = 0.19 \pm 0.10$, $<S(R)> (SbcSm) = 0.35 \pm 0.20$, 
respectively. These values are in good agreement with previous results 
for non--interacting galaxies. 

While $C$ systematically increases from bluer to redder bands, both $A$ and 
$S$ significantly decrease. The $CAS$ parameters present more robust trends 
with the morphological type $T$ and the total $(B-I)$ color in the redder bands, 
suggesting that the basic structure of galaxies is revealed better in the 
IR and NIR bands. The $C$ parameter tends to be higher for earlier--type and redder 
galaxies, while $A$ and $S$ tend to become smaller.
The $A$ parameter could be an excellent way to detect candidates to isolated 
spirals with signs of interaction by means of an automatic analysis of images.  

The loci of our isolated galaxies in the projected planes of the $CAS$ space
depend on the morphological type (and on the color). The major 
difference between the isolated and interacting, starburst and 
ULIR spirals takes place in the $A-S$ plane. 

After the completion of this paper, a work by Taylor-Mager et al. (2007) 
with some aims and results similar to those presented here was posted in the 
arXiv database. Taylor-Mager et al. 
analyzed the $CAS$ parameters for a sample of galaxies (mainly late-types including
peculiars) as a function of wavelength, from UV to IR ($0.15--0.85\mu$m),
with the aim of exploring how the galaxy's appearance changes with rest--frame 
wavelength. The result leads to a measure of the morphological k--correction
for high--redshift galaxies. Their results complement well the ones 
presented here and both are in qualitative agreement, where comparison is possible.
The building of well--defined samples of local isolated galaxies with uniform
and detailed photometric information is of great relevance because 
it provides a fair database for comparison with model predictions
as well as with observed samples of galaxies in other environments
and at higher redshifts. In this paper we present a first step in 
the building of such a sample and discuss some preliminary 
results.

\begin{acknowledgements}

H.M.H.T. thanks the staff of the Observatorio Astron\'omico Nacional in San Pedro 
M\'artir, B. C. for the help with the observation runs. We are grateful to the referee, 
Dr. Michael S. Vogeley, for his comments that improved the presentation of the paper. 
This work was funded by CONACYT grant 42810 to H.M.H.T.
This research has made use of the NASA/IPAC Extragalactic Database (NED) which is 
operated by the Jet Propulsion Laboratory, California Institute of Technology, 
under contract with the National Aeronautics and Space Administration. 
This publication makes use of data products from the Two Micron All Sky Survey, which 
is a joint project of the University of Massachusetts and the Infrared Processing and Analysis 
Center/California Institute of Technology, funded by the National Aeronautics and Space 
Administration and the National Science Foundation.
We acknowledge the usage of the HyperLeda database (http://leda.univ-lyon1.fr).
\end{acknowledgements}


\begin{appendix}
\section{Aperture Photometry}

Since the birth of galaxy photometry (Whitford (1936), 
the amount of photometric data has 
increased exponentially (Prugniel 1987). 
However, these data are inhomogeneous both in quality and 
format: photographic, photoelectric or more recently, CCD 
observations. The data are usually presented as centered 
aperture photometry through circular or elliptical apertures 
or as photometric profiles. In order to take into account 
the continuously growing amount of photometric data and 
at the same time, to make different photometric data 
reports somehow comparable, we present in Table A1
  our estimations of integrated magnitudes 
in two additional concentric circular apertures. Columns (2) and (3) 
give the logarithm of the aperture radius (in units of 0.1 arcmin 
see HyperLeda convention) for each isolated spiral galaxy. 
Columns (4)-(11) give their 
corresponding magnitudes in $B$, $V$, $R$ and $I$ bands, 
respectively. The contribution of the sky to the errors 
in the magnitudes is relatively small at these apertures. 
Typical uncertainties in the magnitudes are 0.11, 0.12, 0.11 
and 0.12 in $B$, $V$, $R$ and $I$ bands, respectively.
 
\placetable{tbl-A1}
 
\end{appendix} 

\twocolumn


\begin{deluxetable}{cccccccccc} 
\tablecolumns{8}
\tablewidth{0pc} 
\tabletypesize{\small}
\tablecaption{Journal of Observations. The number of frames per
filter, the integration time (in seconds), and the mean FWHM for 
each observation (in arcsec) are given.\label{tbl-1}}
\tablehead{
\colhead{} CIG & $B$ & $<B>_{\rm FWHM}$ & $V$ & 
 $<V>_{\rm FWHM}$ & $R$ & $<R>_{\rm FWHM}$ & $I$ 
& $<I>_{\rm FWHM}$}  
\startdata

CIG 1  & 1$\times$1200 &2.3& 1$\times$600  &2.4& 1$\times$300   &2.1& 1$\times$300
&2.4  \\
CIG 4  & 1$\times$1200 &2.3& 1$\times$600  &2.2& 1$\times$300  &2.0& 2$\times$150
&1.8  \\
CIG 33  & 1$\times$1200 &2.3& 1$\times$600  &2.2& 1$\times$300  &2.3& 1$\times$300
&2.5  \\
CIG 53  & 2$\times$1200 &2.1& 2$\times$600 &2.3& 2$\times$300 &2.0& 2$\times$300
&1.8  \\
CIG 56  & 1$\times$1200 &2.4& 1$\times$600  &2.5& 1$\times$300  &2.3& 1$\times$300
&2.2  \\
CIG 68 & 1$\times$1200 &2.1& 2$\times$300 &2.0& 2$\times$120  &2.0& 2$\times$60
&2.3  \\
CIG 80 & 1$\times$1200 &2.7& 2$\times$300  &2.5& 1$\times$300  &2.6& 2$\times$120
&1.8  \\
CIG 103 & 2$\times$1200 &2.0& 1$\times$600 &1.7& 1$\times$300  &1.8& 1$\times$150
&1.4  \\
CIG 116 & 1$\times$1200 &2.1& 1$\times$600  &1.9& 2$\times$150   &1.6& 2$\times$150
&1.6  \\
CIG 123 & 1$\times$1200 &2.0& 1$\times$600 &2.3& 2$\times$150  &1.9& 2$\times$150
&2.0  \\
CIG 138 & 2$\times$1200 &2.2& 2$\times$600  &1.9& 3$\times$150  &1.8& 3$\times$150
&2.1  \\
CIG 139 & 2$\times$1200 &2.1& 1$\times$600 &2.0& 1$\times$300  &1.8& 2$\times$150
&1.6  \\
CIG 144 & 2$\times$1200 &3.0& 1$\times$600  &3.0& 1$\times$300  &2.8& 1$\times$300
&2.6  \\
CIG 151 & 2$\times$1200 &2.1& 2$\times$900 &1.7& 2$\times$300  &1.8& 2$\times$300
&1.9  \\
CIG 154 & 2$\times$1200 &1.9& 1$\times$600 &1.5& 1$\times$300  &1.5& 1$\times$300
&1.5  \\
CIG 168 & 1$\times$1200 &1.7& 1$\times$600 &1.5& 1$\times$300  &1.5& 1$\times$300
&1.5  \\
CIG 175 & 1$\times$1200 &1.8& 1$\times$600 &1.6& 1$\times$300  &1.7& 1$\times$300
&1.7  \\
CIG 180 & 2$\times$1200 &1.9& 2$\times$600  &2.0& 1$\times$300  &1.7& 4$\times$150
&1.8  \\
CIG 188 & 2$\times$1200 &1.8& 1$\times$900  &1.8& 2$\times$300  &1.7& 1$\times$300
&1.8  \\
CIG 208 & 1$\times$1200 &1.8& 1$\times$900  &1.7& 1$\times$480  &1.7& 1$\times$480
&1.8  \\
CIG 213 & 1$\times$1200 &1.8& 1$\times$600  &1.7& 1$\times$300  &1.7& 1$\times$300
&1.8  \\
CIG 224 & 1$\times$1200 &1.7& 2$\times$600  &1.7& 1$\times$300  &1.8& 1$\times$240
&1.8  \\
CIG 237 & 1$\times$1200 &1.8& 1$\times$900  &1.7& 1$\times$300  &1.7& 1$\times$300
&1.8  \\
CIG 309 & 2$\times$600 &1.8& 2$\times$300  &1.7& 2$\times$60  &1.7& 2$\times$60
&1.7  \\
CIG 314 & 1$\times$1200 &1.8& 1$\times$600  &1.7& 1$\times$300  &1.8& 1$\times$300
&1.8  \\
CIG 434 & 1$\times$1200 &1.8& 1$\times$900  &1.7& 2$\times$300  &1.7& 1$\times$600
&1.7  \\
CIG 472 & 1$\times$600 &1.8& 1$\times$600  &1.7& 1$\times$300  &1.7& 1$\times$300
&1.7  \\
CIG 518 & 1$\times$600 &1.8& 1$\times$600  &1.8& 1$\times$300  &1.8& 1$\times$300
&1.7  \\
CIG 528 & 1$\times$1200 &1.9& 1$\times$600  &1.8& 1$\times$300  &1.8& 1$\times$300
&1.9  \\
CIG 549 & 1$\times$600 &1.9& 1$\times$300  &1.8& 2$\times$120  &1.7& 1$\times$120
&1.9  \\
CIG 604 & 1$\times$900 &1.8& 1$\times$600  &1.8& 2$\times$120  &1.8& 1$\times$120
&1.7  \\
CIG 605 & 1$\times$900 &1.9& 1$\times$480  &2.0& 2$\times$180  &2.0& 1$\times$180
&1.9  \\
CIG 691 & 1$\times$900 &1.9& 1$\times$600  &2.0& 1$\times$300  &2.0& 1$\times$300
&1.9  \\
CIG 710 & 1$\times$600 &1.9& 1$\times$600  &2.0& 1$\times$300  &2.0& 1$\times$300
&1.9  \\
CIG 889 & 1$\times$1200 &2.0& 1$\times$600  &2.1& 1$\times$300  &2.0& 2$\times$150
&1.9  \\
CIG 906 & 2$\times$1200 &2.3& 1$\times$600  &2.7& 2$\times$300  &2.0& 2$\times$300
&1.8  \\
CIG 910 & 2$\times$1200 &2.3& 1$\times$600  &2.0& 1$\times$300   &1.9& 1$\times$300
&1.8  \\
CIG 911 & 2$\times$1200 &1.6& 2$\times$600  &1.6& 1$\times$300  &1.5& 1$\times$300
&1.6  \\
CIG 935 & 2$\times$1200 &1.6& 2$\times$600  &1.6& 2$\times$300  &1.5& 2$\times$300
&1.6  \\
CIG 976 & 2$\times$1200 &1.7& 2$\times$600  &1.8& 2$\times$300  &1.8& 2$\times$300
&1.9  \\
CIG 983 & 2$\times$1200 &1.6& 1$\times$600  &1.6& 1$\times$300  &1.5& 2$\times$300
&1.6  \\
CIG 1004 & 2$\times$1200 &1.6& 1$\times$600  &1.6& 1$\times$300  &1.5& 1$\times$300
&1.6  \\
CIG 1009 & 2$\times$1200 &1.8& 1$\times$600  &1.6& 1$\times$300  &1.9& 2$\times$150
&1.9  \\
CIG 1023 & 1$\times$1200 &1.5& 1$\times$600  &1.6& 1$\times$300  &1.6& 3$\times$120
&1.6  \\
\enddata
\end{deluxetable}

\begin{deluxetable}{ccccccc} 
\tablecolumns{8}
\tablewidth{0pc} 
\tablecaption{General Data for the Observed Isolated Spiral Galaxies. \label{tbl-2}}
\tablehead{
\colhead{} CIG & Identif & $B$ mag (LEDA)  & Type (LEDA) & 
 $B$ mag (NED)  &  $V_{Rad}$ (km s$^{-1}$)}  
\startdata 

CIG 1  & UGC 00005 & 14.08 & Sbc & 13.97 & 7243 & \\
CIG 4  & NGC 7817 & 12.74 & Sbc & 12.56 & 2391 &  \\
CIG 33  & NGC 0237  & 13.72 & SABc & 13.70 & 4136 & \\
CIG 53  & NGC 0575   & 13.76 & Sc  & 13.45 & 3185 & \\
CIG 56  & NGC 0622 & 14.08 & Sb & 13.71 & 5113 & \\
CIG 68  & NGC 0718 & 12.59 & Sa  & 12.59 & 1683 & \\
CIG 80  & NGC 0772 & 10.29 & Sb & 11.09 & 2473 & \\
CIG 103  & NGC 0918 & 13.07 &  Sc & 13.05 & 1536 & \\
CIG 116  & NGC 1050 & 13.55 & SBa & 13.47 & 3989 & \\
CIG 123  & IC 0302  & 13.60 &  Sbc & 13.59 & 5854 & \\
CIG 138 & UGC 02936 & 14.76 & Sc & 15.00 & 3743 & \\
CIG 139 & NGC 1507 & 12.79 & SBd  & 12.89 & 769 & \\
CIG 144 & UGC 02988 & 15.52 & Sb & 14.90 & 3868  & \\
CIG 151 & UGC 03059 & 15.11 &  Sd  & 14.70 & 4743  & \\
CIG 154 & UGC 03171 & 14.84 & Sc & 14.78 & 4475 &  \\
CIG 168 & IC 2166 & 13.24 & SABc & 13.20 & 2892 &  \\
CIG 175 & UGC 03580 & 13.09 & SABa & 12.71 & 1439 &  \\  
CIG 180 & NGC 2344 & 12.89 & SABb  & 12.81 & 1126 & \\
CIG 188 & UGC 03826 & 14.59 & SABc & 14.10  & 1947 & \\ 
CIG 208 & UGC 04054 & 14.80 & Sb  & 14.30 & 2175 & \\ 
CIG 213 & PGC 022141 & 14.79 & Sa & 14.88 & 6089 & \\
CIG 224 & NGC 2500 & 12.22 & Scd  & 12.20 & 693 & \\
CIG 237 & UGC 04277 & 14.82  & Sc & 14.90 & 5650 & \\ 
CIG 309 & NGC 2775 & 11.14 & Sab & 11.03 & 1347 & \\  
CIG 314 & NGC 2776 & 12.18 & SABc & 12.14 & 2798 & \\  
CIG 434 & UGC 05829 & 13.73 & I & 13.73 & 780 & \\
CIG 472 & NGC 3596 & 11.79 & SABc  & 11.95 & 1267 & \\ 
CIG 518 & NGC 4062 & 11.88 & SABc  & 11.90 & 939  & \\ 
CIG 528 & NGC 4357 & 13.25 & Sbc  & 13.20 & 4367 & \\ 
CIG 549 & NGC 4651 & 11.38 & Sc  & 11.39 & 911 & \\
CIG 604 & NGC 5377 & 12.16 & Sa & 12.24 & 2044 & \\
CIG 605 & NGC 5375 & 12.79 & SBab & 12.40 & 2572 & \\
CIG 691 & NGC 5964 & 13.33 & SBcd  & 12.60 & 1555 & \\  
CIG 710 & NGC 6015 & 11.62 & Sc  & 11.69 & 1111 & \\
CIG 889 & NGC 6969 & 15.33 & Sa  & 14.89 & 4778 & \\
CIG 906 & UGC 11723 & 14.74 & Sb  & 14.70 & 4928 & \\
CIG 910 & IC 5104  & 14.33 & Sab & 14.27 & 5110 & \\
CIG 911 & NGC 7056  & 13.67 & SBbc  & 13.75 & 5501 & \\
CIG 935 & NGC 7156 & 13.29 & SABc & 13.11 & 4023 & \\
CIG 976 & NGC 7328 & 13.93 & Sab & 13.98 & 2886 & \\
CIG 983 & UGC 12173 & 13.56 & SABc & 13.49 & 4960 & \\
CIG 1004 & NGC 7479 & 11.71 & SBbc & 11.60 & 2443  & \\
CIG 1009 & NGC 7514 & 13.55 & Sbc & 13.54 & 5005 & \\
CIG 1023 & UGC 12646 & 14.22 & Sb & 13.99 & 8143 & \\
\enddata
\end{deluxetable}


\begin{deluxetable}{cccccccccc} 
\tablecolumns{8} 
\tablewidth{0pc} 
\tablecaption{Apparent Magnitudes and Color Indices. \label{tbl-3}}
\tablehead{
\colhead{} CIG & Log (A) & $B$ & $V$ & 
$R$ & $I$ & $B-V$ & $B-R$ & $B-I$}  
\startdata 
 
CIG1  & 1.59  & 13.92 &    13.32 &     12.40 &     12.08 &      0.60 &     1.51 &      1.83 & \\
CIG4  & 1.62  & 12.86 &    11.94 &     11.30 &     10.27 &      0.92 &     1.56 &      2.59 & \\
CIG33 & 1.53  & 13.65 &    13.06 &     12.53 &     11.85 &      0.59 &     1.12 &      1.80 & \\
CIG53 & 1.53  & 13.65 &    13.02 &     12.51 &     11.81 &      0.63 &     1.14 &      1.83 & \\
CIG56 & 1.59  & 13.96 &    13.30 &     12.71 &     12.00 &      0.66 &     1.26 &      1.96 & \\
CIG68 & 1.58  & 12.52 &    11.67 &     11.10 &     10.29 &      0.84 &     1.42 &      2.23 & \\
CIG80 & 1.59  & 11.65 &    10.80 &     10.23 &      9.38 &      0.85 &     1.42 &      2.27 & \\
CIG103& 1.58  & 13.28 &    12.37 &     11.68 &     10.75 &      0.92 &     1.60 &      2.53 & \\
CIG116& 1.55  & 13.71 &    12.81 &     12.25 &     11.48 &      0.90 &     1.46 &      2.23 & \\
CIG123& 1.59  & 13.81 &    13.00 &     12.33 &     11.60 &      0.81 &     1.47 &      2.20 & \\ 
CIG138& 1.59  & 14.60 &    13.33 &     12.51 &     11.28 &      1.27 &     2.09 &      3.31 & \\
CIG139& 1.60  & 12.89 &    12.36 &     11.90 &     11.45 &      0.52 &     0.98 &      1.43 & \\
CIG144& 1.59  & 15.31 &    13.95 &     13.03 &     11.63 &      1.36 &     2.28 &      3.68 & \\
CIG151& 1.57  & 14.96 &    14.07 &     13.34 &     12.37 &      0.89 &     1.62 &      2.59 & \\
CIG154& 1.52  & 14.62 &    14.05 &     13.48 &     12.93 &      0.58 &     1.15 &      1.69 & \\
CIG168& 1.50  & 12.68 &    12.01 &     11.45 &     10.86 &      0.66 &     1.22 &      1.82 & \\
CIG175& 1.50  & 13.01 &    12.57 &     12.08 &     11.50 &      0.44 &     0.92 &      1.51 & \\
CIG180& 1.53  & 12.99 &    12.14 &     11.57 &     10.80 &      0.85 &     1.41 &      2.19 & \\
CIG188& 1.50  & 13.20 &    12.70 &     12.25 &     11.99 &      0.50 &     0.94 &      1.20 & \\
CIG208& 1.50  & 14.58 &    14.01 &     13.55 &     12.86 &      0.58 &     1.03 &      1.72 & \\
CIG213& 1.55  & 14.74 &    13.80 &     00.00 &     00.00 &      0.94 &     0.00 &      0.00 & \\
CIG224& 1.50  & 12.06 &    11.62 &     11.34 &     10.85 &      0.44 &     0.72 &      1.20 & \\
CIG237& 1.50  & 14.39 &    13.53 &     12.95 &     12.27 &      0.86 &     1.44 &      2.11 & \\
CIG309& 1.50  & 11.31 &    10.39 &      9.72 &      9.09 &      0.91 &     1.58 &      2.30 & \\
CIG314& 1.50  & 12.26 &    11.69 &     11.26 &     10.77 &      0.56 &     0.99 &      1.49 & \\
CIG434& 1.50  & 13.31 &    13.09 &     12.84 &     12.85 &      0.21 &     0.46 &      0.45 & \\
CIG472& 1.50  & 12.01 &    11.49 &     10.99 &     10.45 &      0.51 &     1.01 &      1.55 & \\
CIG518& 1.50  & 11.90 &    11.21 &     10.62 &     10.03 &      0.69 &     1.28 &      1.90 & \\
CIG528& 1.50  & 13.36 &    12.68 &     12.14 &     11.84 &      0.67 &     1.21 &      1.51 & \\
CIG549& 1.50  & 11.65 &    10.92 &     10.42 &      9.84 &      0.73 &     1.22 &      1.80 & \\
CIG604& 1.50  & 12.39 &    11.46 &     10.87 &     10.25 &      0.92 &     1.51 &      2.13 & \\
CIG605& 1.50  & 12.78 &    12.08 &     11.65 &     11.65 &      0.70 &     1.13 &      1.14 & \\
CIG691& 1.50  & 12.45 &    11.82 &     11.48 &     10.89 &      0.63 &     0.97 &      1.56 & \\
CIG710& 1.50  & 11.69 &    11.14 &     10.61 &     10.08 &      0.56 &     1.08 &      1.61 & \\
CIG889& 1.57  & 14.47 &    13.52 &     12.84 &     12.07 &      0.95 &     1.63 &      2.40 & \\
CIG906& 1.43  & 14.89 &    14.09 &     13.44 &     12.61 &      0.81 &     1.46 &      2.29 & \\
CIG910& 1.43  & 14.81 &    13.82 &     13.25 &     12.49 &      0.99 &     1.57 &      2.32 & \\
CIG911& 1.58  & 13.86 &    13.09 &     12.48 &     11.75 &      0.77 &     1.38 &      2.11 & \\
CIG935& 1.56  & 13.41 &    12.72 &     12.24 &     11.53 &      0.69 &     1.17 &      1.88 & \\
CIG976& 1.53  & 13.99 &    13.10 &     12.55 &     11.46 &      0.89 &     1.44 &      2.53 & \\
CIG983& 1.57  & 13.81 &    13.03 &     12.28 &     11.64 &      0.78 &     1.53 &      2.18 & \\
CIG1004& 1.57 & 11.82 &    11.04 &     10.47 &      9.66 &      0.78 &     1.35 &      2.16 & \\
CIG1009& 1.35 & 13.98 &    13.15 &     12.55 &     11.80 &      0.84 &     1.44 &      2.18 & \\
CIG1023& 1.54 & 14.08 &    13.26 &     12.71 &     11.73 &      0.82 &     1.37 &      2.35 & \\
\enddata
\end{deluxetable}

\begin{deluxetable}{ccccccccc} 
\tablecolumns{8} 
\tablewidth{0pc} 
\tablecaption{Corrected Colors and Absolute Magnitudes. \label{tbl-4}}
\tablehead{
\colhead{} CIG &  $(B-V)_c$  &  $(B-R)_c$  &  $(B-I)_c$  & 
$M_{B}$ & $M_{V}$ &  $M_{R}$ &  $M_{I}$ }  
\startdata

CIG1   & 0.36  &  1.28  &    1.39 &   -22.11  &  -22.47 &   -23.39  &  -23.50  \\
CIG4   & 0.65  &  1.27  &    2.08 &   -21.11  &  -21.75 &   -22.37  &  -23.18  \\
CIG33  & 0.47  &  1.00  &    1.59 &   -20.68  &  -21.15 &   -21.67  &  -22.26  \\
CIG53  & 0.54  &  1.01  &    1.64 &   -20.03  &  -20.57 &   -21.04  &  -21.67  \\
CIG56  & 0.54  &  1.12  &    1.73 &   -20.84  &  -21.39 &   -21.96  &  -22.58  \\
CIG68  & 0.79  &  1.34  &    2.10 &   -19.65  &  -20.44 &   -20.99  &  -21.75  \\
CIG80  & 0.64  &  1.15  &    1.84 &   -22.00  &  -22.64 &   -23.16  &  -23.85  \\
CIG103 & 0.46  &  0.93  &    1.53 &   -20.33  &  -20.79 &   -21.26  &  -21.86  \\
CIG116 & 0.78  &  1.31  &    1.98 &   -20.60  &  -21.38 &   -21.90  &  -22.57  \\
CIG123 & 0.54  &  1.10  &    1.62 &   -21.95  &  -22.49 &   -23.04  &  -23.57  \\
CIG138 & 0.54  &  1.06  &    1.75 &   -22.23  &  -22.77 &   -23.29  &  -23.98  \\
CIG139 & 0.19  &  0.57  &    0.75 &   -18.59  &  -18.78 &   -19.16  &  -19.33  \\ 
CIG144 & 0.53  &  1.11  &    1.89 &   -22.05  &  -22.58 &   -23.16  &  -23.95  \\
CIG151 & 0.38  &  0.93  &    1.53 &   -21.34  &  -21.72 &   -22.28  &  -22.87  \\
CIG154 & 0.43  &  0.93  &    1.37 &   -20.01  &  -20.44 &   -20.94  &  -21.38  \\
CIG168 & 0.44  &  0.93  &    1.35 &   -21.33  &  -21.77 &   -22.26  &  -22.68  \\
CIG175 & 0.30  &  0.75  &    1.23 &   -19.11  &  -19.41 &   -19.86  &  -20.34  \\
CIG180 & 0.74  &  1.25  &    1.94 &   -18.52  &  -19.26 &   -19.77  &  -20.46  \\
CIG188 & 0.40  &  0.80  &    0.97 &   -19.45  &  -19.84 &   -20.25  &  -20.42  \\ 
CIG208 & 0.34  &  0.77  &    1.28 &   -18.83  &  -19.17 &   -19.60  &  -20.10  \\ 
CIG213 & 0.86  &  0.00  &    0.00 &   -20.30  &  -21.17 &    00.00  &   00.00  \\
CIG224 & 0.40  &  0.65  &    1.10 &   -18.12  &  -18.52 &   -18.77  &  -19.22  \\ 
CIG237 & 0.46  &  1.02  &    1.32 &   -21.99  &  -22.45 &   -23.01  &  -23.31  \\
CIG309 & 0.83  &  1.48  &    2.12 &   -20.49  &  -21.32 &   -21.97  &  -22.61  \\
CIG314 & 0.42  &  0.83  &    1.21 &   -21.33  &  -21.75 &   -22.16  &  -22.54 \\
CIG434 & 0.18  &  0.42  &    0.39 &   -17.05  &  -17.23 &   -17.47  &  -17.44 \\ 
CIG472 & 0.48  &  0.97  &    1.48 &   -19.43  &  -19.91 &   -20.39  &  -20.90 \\ 
CIG518 & 0.56  &  1.16  &    1.65 &   -19.32  &  -19.88 &   -20.48  &  -20.97 \\
CIG528 & 0.44  &  0.96  &    1.05 &   -21.59  &  -22.03 &   -22.55  &  -22.63 \\ 
CIG549 & 0.65  &  1.14  &    1.64 &   -19.28  &  -19.93 &   -20.42  &  -20.92 \\ 
CIG604 & 0.79  &  1.38  &    1.86 &   -20.59  &  -21.39 &   -21.98  &  -22.45 \\ 
CIG605 & 0.64  &  1.05  &    0.98 &   -20.34  &  -20.98 &   -21.39  &  -21.32 \\ 
CIG691 & 0.53  &  0.83  &    1.34 &   -19.71  &  -20.24 &   -20.54  &  -21.05 \\ 
CIG710 & 0.41  &  0.94  &    1.33 &   -19.91  &  -20.32 &   -20.85  &  -21.24 \\ 
CIG889 & 0.66  &  1.30  &    1.81 &   -21.05  &  -21.71 &   -22.35  &  -22.86 \\ 
CIG906 & 0.50  &  1.16  &    1.72 &   -20.73  &  -21.22 &   -21.88  &  -22.45 \\ 
CIG910 & 0.68  &  1.17  &    1.65 &   -20.97  &  -21.64 &   -22.14  &  -22.62 \\ 
CIG911 & 0.66  &  1.21  &    1.86 &   -21.08  &  -21.74 &   -22.29  &  -22.94 \\ 
CIG935 & 0.59  &  1.03  &    1.67 &   -20.81  &  -21.40 &   -21.84  &  -22.47 \\ 
CIG976 & 0.67  &  1.17  &    2.09 &   -20.12  &  -20.79 &   -21.30  &  -22.22 \\ 
CIG983 & 0.50  &  1.17  &    1.59 &   -21.63  &  -22.12 &   -22.80  &  -23.21 \\ 
CIG1004&  0.58 &  1.08  &    1.74 &   -21.72  &  -22.30 &   -22.80  &  -23.46 \\ 
CIG1009&  0.67 &  1.22  &    1.84 &   -20.98  &  -21.65 &   -22.20  &  -22.82 \\
CIG1023&  0.68 &  1.20  &    2.08 &   -21.85  &  -22.53 &   -23.04  &  -23.92 \\    
\enddata
\end{deluxetable} 


\begin{deluxetable}{ccccccc} 
\tablecolumns{8} 
\tablewidth{0pc} 
\tablecaption{Final Morphological Classification. \label{tbl-5}}
\tablehead{
\colhead{} CIG & Type (NED)  & Type (This work) &  Bars/Rings  & 
 Optical Arms &  Bar Ellipticity $\epsilon_{\rm max}$ }  
\startdata

CIG1   & SABbc      & SBbc     & B   & multi & 0.24  \\ 
CIG4   & SAbc       & SABc     & B   & multi & --- \\
CIG33  & SAB(rs)cd  & SB(rs)c  & B/R &       & 0.27  \\
CIG53  & SB(rs)c    & SB(r)bc  & B/R & multi & 0.59   \\
CIG56  & SB(rs)b    & SB(r)b   & B/R &       & 0.38  \\
CIG68  & SAB(s)a    & SB(r)a   & B/R &       & 0.39  \\
CIG80  & SA(s)b     & SB(r)b   & B/R & multi & 0.20  \\
CIG103 & SAB(rs)c   & SB(r)c   & B/R & multi & 0.19  \\
CIG116 & (R)SB(s)a  & RSB(s)a  & B   &       & 0.42  \\
CIG123 & SB(rs)bc   & SB(r)c   & B/R & multi & 0.46  \\
CIG138 & SB(s)d     & SBc      & B   & multi & 0 \\ 
CIG139 & SB(s)m pec & SB(s)m   & B   &       &   \\
CIG144 & Sb         & SABb     & B(peanut?)  &  & --- \\
CIG151 & SAdm       & SABc     & B   & multi & ---  \\
CIG154 & SBcd       & SB(r)cd  & B/R & multi & 0.36  \\
CIG168 & SAB(s)bc   & SAB(s)cd &     & multi & --- \\ 
CIG175 & SA(s)a pec & Sa pec   &     &       &  \\
CIG180 & SA(rs)c    & SA(r)b   & R   & multi &   \\
CIG188 & SAB(s)d    & SB(s)d   & B   & multi &   \\
CIG208 & Sb         & SABcd    & B   &       & ---   \\
CIG213 & S0         & RSB0     & B/R &       & 0.39 \\
CIG224 & SB(rs)d    & RSBd     & B/R & multi & 0.40  \\
CIG237 & Sc         & Sc       &     &       &   \\
CIG309 & SA(r)ab    & RS(r)a   & R   &       &   \\ 
CIG314 & SAB(rs)c   & SAB(rs)c & B/R & multi & ---   \\ 
CIG434 & Im         & Im       & B   &       & 0.80   \\
CIG472 & SAB(rs)c   & SA(rs)c  & R   & multi &    \\
CIG518 & SA(s)c     & SB(s)bc  & B   & multi & 0.20   \\  
CIG528 & SAbc       & SA(rs)cd & R   & multi &    \\
CIG549 & SA(rs)c    & SB(rs)c  & B/R & multi & 0.31   \\
CIG604 & (R)SB(s)a  & RSB(s)a pec & B(boxy)/R  &   & 0.22 \\
CIG605 & SB(r)ab    & SB(r)b   & B/R & multi & 0.44   \\
CIG691 & SB(rs)d    & SB(rs)cd & B/R & multi & 0.53   \\
CIG710 & SA(s)cd    & SB(s)cd  & B   & multi & 0.20   \\
CIG889 & Sa         & SBa      & B(peanut)   &  & --- \\
CIG906 & Sb         & Sb       &     &       &  \\
CIG910 & SBab       & SBab     & B   &       & --- \\
CIG911 & SBb        & SB(r)b   & B/R & multi & 0.47 \\
CIG935 & SAB(rs)cd  & SAB(rs)c & B/R & multi & --- \\
CIG976 & Sab        & SA(s)ab  &     & multi &  \\
CIG983 & SAB(rs)c   & SB(rs)cd & B/R& multi & 0.30 \\ 
CIG1004 & SB(s)c    & SB(s)c   & B   & multi & 0.60 \\ 
CIG1009 & Sa        & S(r)ab   & R   & multi &  \\
CIG1023 & SB(r)b    & RSB(r)b  & B/R &       & 0.40 \\    
\enddata
\end{deluxetable}

\begin{deluxetable}{ccccccc} 
\tablecolumns{8} 
\tablewidth{0pc} 
\tablecaption{$R$ Band $CAS$ Parameters for Isolated Spiral Galaxies. \label{tbl-6}}
\tablehead{
\colhead{} CIG & Inclination &  $C(R)$  &  $A(R)$ & $S(R)$ & }  
\startdata
  
CIG1     & 66.8 & 2.88$\pm$0.09&  0.19$\pm$0.02&  0.30$\pm$0.07& \\
CIG4     & 84.1 & 3.30$\pm$0.05&  0.15$\pm$0.02&  0.40$\pm$0.08& \\
CIG33    & 54.1 & 3.13$\pm$0.12&  0.09$\pm$0.21&  0.16$\pm$0.03& \\
CIG53    & 34.6 & 2.57$\pm$0.05&  0.08$\pm$0.02&  0.31$\pm$0.01& \\
CIG56    & 51.4 & 4.19$\pm$0.15&  0.06$\pm$0.01&  0.18$\pm$0.03& \\
CIG68    & 32.1 & 4.85$\pm$0.11&  0.06$\pm$0.01&  0.11$\pm$0.02& \\
CIG80    & 48.5 & 3.66$\pm$0.04&  0.13$\pm$0.01&  0.01$\pm$0.01& \\
CIG103   & 57.5 & 2.84$\pm$0.03&  0.12$\pm$0.03&  0.48$\pm$0.01& \\
CIG116   & 47.0 & 3.62$\pm$0.12&  0.14$\pm$0.07&  0.20$\pm$0.04& \\
CIG123   & 49.7 & 3.18$\pm$0.07&  0.17$\pm$0.02&  0.60$\pm$0.02& \\
CIG138   & 78.7 & 3.65$\pm$0.04&  0.05$\pm$0.03&  0.12$\pm$0.09& \\
CIG139   & 90.0 & 3.30$\pm$0.04&  0.23$\pm$0.02&  0.40$\pm$0.01& \\
CIG144   & 90.0 & 4.32$\pm$0.12&  0.10$\pm$0.03&  0.27$\pm$0.03& \\
CIG151   & 90.0 & 3.17$\pm$0.08&  0.23$\pm$0.04&  0.56$\pm$0.02& \\
CIG154   & 37.0 & 2.90$\pm$0.10&  0.07$\pm$0.03&  0.07$\pm$0.03& \\
CIG168   & 57.1 & 2.71$\pm$0.06&  0.49$\pm$0.01&  0.70$\pm$0.09& \\
CIG175   & 63.0 & 3.94$\pm$0.17&  0.11$\pm$0.02&  0.16$\pm$0.04& \\
CIG180   & 26.0 & 3.69$\pm$0.07&  0.01$\pm$0.02&  0.05$\pm$0.05& \\
CIG188   & 36.5 & 2.62$\pm$0.04&  0.41$\pm$0.02&  0.65$\pm$0.02& \\
CIG208   & 90.0 & 3.83$\pm$0.16&  0.08$\pm$0.03&  0.33$\pm$0.01& \\
CIG224   & 21.7 & 2.89$\pm$0.04&  0.18$\pm$0.03&  0.36$\pm$0.01& \\
CIG237   & 90.0 & 3.90$\pm$0.13&  0.19$\pm$0.08&  0.45$\pm$0.02& \\
CIG309   & 40.6 & 4.13$\pm$0.08&  0.01$\pm$0.08&  0.14$\pm$0.05& \\
CIG314   & 58.3 & 3.42$\pm$0.09&  0.18$\pm$0.02&  0.53$\pm$0.08& \\
CIG434   & 24.5 & 2.81$\pm$0.03&  0.42$\pm$0.07&  0.65$\pm$0.04& \\
CIG472   & 23.2 & 2.96$\pm$0.08&  0.15$\pm$0.01&  0.51$\pm$0.07& \\
CIG518   & 68.2 & 2.97$\pm$0.06&  0.14$\pm$0.02&  0.53$\pm$0.07& \\
CIG528   & 73.6 & 3.49$\pm$0.13&  0.15$\pm$0.02&  0.35$\pm$0.01& \\
CIG549   & 53.0 & 3.58$\pm$0.09&  0.11$\pm$0.03&  0.06$\pm$0.03& \\
CIG604   & 70.6 & 4.73$\pm$0.12&  0.01$\pm$0.01&  0.23$\pm$0.05& \\
CIG605   & 42.8 & 4.62$\pm$0.12&  0.07$\pm$0.03&  0.53$\pm$0.01& \\
CIG691   & 40.7 & 2.51$\pm$0.04&  0.43$\pm$0.02&  0.65$\pm$0.01& \\
CIG710   & 67.4 & 3.22$\pm$0.05&  0.24$\pm$0.01&  0.50$\pm$0.08& \\
CIG889   & 90.0 & 4.09$\pm$0.21&  0.01$\pm$0.06&  0.05$\pm$0.02& \\
CIG906   & 90.0 & 4.09$\pm$0.07&  0.19$\pm$0.03&  0.40$\pm$0.02& \\
CIG910   & 90.0 & 3.75$\pm$0.12&  0.08$\pm$0.02&  0.30$\pm$0.08& \\
CIG911   & 20.9 & 3.27$\pm$0.12&  0.08$\pm$0.08&  0.11$\pm$0.03& \\
CIG935   & 40.0 & 3.03$\pm$0.08&  0.12$\pm$0.03&  0.19$\pm$0.05& \\
CIG976   & 78.3 & 3.26$\pm$0.10&  0.15$\pm$0.02&  0.41$\pm$0.08& \\
CIG983   & 62.0 & 2.73$\pm$0.06&  0.15$\pm$0.02&  0.17$\pm$0.05& \\
CIG1004  & 36.4 & 3.31$\pm$0.03&  0.23$\pm$0.08&  0.45$\pm$0.07& \\
CIG1009  & 48.7 & 3.16$\pm$0.10&  0.05$\pm$0.04&  0.14$\pm$0.04& \\
CIG1023  & 49.7 & 4.39$\pm$0.17&  0.05$\pm$0.01&  0.20$\pm$0.04& \\
\enddata
\end{deluxetable}


\begin{deluxetable}{cccccccccccc} 
\tablecolumns{11}
\tablewidth{0pc} 
\tablecaption{Magnitudes at Different Circular Apertures. \label{tbl-A1}}
\tablehead{
\colhead{} CIG &  $Log A1$ &  $B$  &  $V$  &  $R$  
&  $I$  &  $Log A2$  &  $B$  &  $V$  &  $R$  &  $I$ }
\startdata 

CIG1 &1.11&14.10&13.41&12.81&12.22 &1.41&14.01&13.39&12.65&12.20& \\
CIG4 &1.14&13.24&12.31&11.65&10.75 &1.44&12.89&11.97&11.33&10.35&  \\
CIG33 &1.05& 13.82&13.19&12.67&12.03 &1.35& 13.67&13.06&12.54&11.87& \\
CIG53 &1.10& 14.03&13.34&12.83&12.16 &1.35& 13.68&13.02&12.52&11.83& \\
CIG56 &1.11& 14.07&13.37&12.80&12.11 &1.41& 13.98&13.31&12.73&12.03& \\
CIG68 &1.10& 12.87&12.03&11.48&10.75 &1.40& 12.55&11.72&11.16&10.38& \\
CIG80 &1.12& 12.48&11.57&10.96&10.13 &1.42& 11.87&11.01&10.41&9.57& \\
CIG103 &1.10& 14.07&13.12&12.44&11.55&1.40& 13.40&12.48&11.80&10.91& \\
CIG116 &1.08& 13.91&13.04&12.48&11.64 &1.38& 13.72&12.83&12.27&11.48& \\
CIG123 &1.11& 14.22&13.31&12.67&11.85 &1.41& 13.84&13.01&12.37&11.61& \\
CIG138 &1.11& 15.02&13.80&12.90&11.77 &1.41& 14.62&13.40&12.53&11.35& \\
CIG139 &1.13& 13.34&12.80&12.35&11.76 &1.43& 12.93&12.38&11.93&11.45& \\
CIG144 &1.12& 15.53&14.18&13.26&12.08 &1.42& 15.31&13.996&13.06&11.79& \\
CIG151 &1.10& 15.24&14.32&13.65&12.85 &1.40& 15.00&14.16&13.49&12.81& \\
CIG154 &1.05& 14.71&14.09&13.59&12.99 &1.35& 14.63&14.05&13.52&12.97& \\
CIG168 &1.20& 13.272 & 12.571 & 11.960 & 11.131&1.50   & 12.623 & 11.980 & 11.386 & 10.536& \\
CIG175 &1.20& 13.418 & 12.861 & 12.361 & 11.668&1.50    & 13.116 &  12.657 &  12.195 & 11.346& \\
CIG180 &1.06& 13.50&12.61&12.03&11.27&1.36& 13.11&12.24&11.67&10.91& \\
CIG188 &1.20& 13.803  &   13.309  &   12.898  &   12.176&1.50    & 13.137  &    12.68  &   12.255  &   11.475& \\
CIG208 &1.20& 14.62  &   14.007  &   13.593  &   12.935 &1.50    & 14.275  &   13.851  &   13.478  &   12.772& \\
CIG224 &1.20& 12.73  &   12.212  &   11.775  &   11.085 &1.50    & 12.025  &   11.501  &   11.127  &   10.098& \\
CIG237 &1.20& 14.825  &   13.849   &   13.17  &   12.279&1.50    & 14.341  &   13.522  &   12.936  &   11.928& \\
CIG309 &1.20& 11.765  &   10.831  &   10.164   &    9.35&1.50    & 11.377 &    10.445  &    9.778  &    8.927& \\
CIG314 &1.20& 12.586  &      12.0   &  11.527  &   10.909&1.50    & 12.244  &   11.711 &    11.245  &   10.626& \\
CIG434 &1.20& 14.036  &   13.787  &   13.417  &   13.024&1.50    & 13.292  &   13.145  &   12.809  &   12.394& \\
CIG472 &1.20& 12.4  &   11.822  &   11.282  &    10.617& 1.50   & 12.098  &   11.567  &   11.042 &    10.365& \\
CIG518 &1.20& 12.418  &   11.677  &    11.07  &   10.332&1.50    & 11.962  &   11.257 &    10.665   &   9.918& \\
CIG528 &1.20& 13.571   &   12.79  &   12.233  &   11.548&1.50    & 13.336  &   12.645  &   12.135 &    11.463& \\
CIG549 &1.20& 12.002  &   11.235  &   10.676  &    9.978&1.50    & 11.621  &    10.91  &   10.381  &    9.686& \\
CIG604 &1.20& 12.731  &   11.844  &   11.196  &   10.426&1.50    & 12.386  &   11.551  &   10.917  &    10.14& \\
CIG605 &1.20& 13.207  &   12.337  &    11.94  &   11.169&1.50    & 12.778  &   11.833  &   11.555   &  10.446& \\
CIG691 &1.20& 13.273  &   12.609  &   12.095 &    11.474&1.50    & 12.404 &    11.819  &   11.363  &   10.766& \\
CIG710 &1.20& 12.212    &  11.62   &  11.078   &  10.399&1.50    & 11.733  &   11.208  &   10.682  &   10.044& \\
CIG889 &1.10& 14.52&13.53&12.90&12.11&1.40& 14.50&13.53&12.88&12.08& \\
CIG906 &0.96& 15.26&14.33&13.66&12.80&1.26& 14.95&14.09&13.44&12.61& \\
CIG910 &0.96& 14.99&14.08&13.44&12.65&1.26& 14.82&13.89&13.28&12.52& \\
CIG911 &1.10& 13.88&13.16&12.53&11.83&1.40& 13.86&13.12&12.49&11.76& \\
CIG935 &1.08& 13.561  &   12.866  &   12.375  &   11.716&1.39& 13.424  &   12.731  &   12.245 &    11.558& \\
CIG976 &1.06& 14.102  &    13.29   &  12.747  &   11.992&1.36& 13.962  &   13.157   &  12.623  &   11.852& \\
CIG983 &1.10& 14.061   &  13.265  &   12.626  &   11.921&1.40& 13.825  &   13.051  &   12.336   &  11.681& \\
CIG1004 &1.10& 12.916  &   12.019   &  11.402   &  10.583&1.40& 12.087   &   11.27  &   10.687   &    9.88& \\
CIG1009 &0.88& 14.276   &  13.421  &   12.853  &   12.135&1.18& 13.989  &    13.15  &   12.572  &   11.831& \\
CIG1023 &1.09& 14.388  &   13.511  &    12.94  &    12.15&1.39&  14.124 &    13.279  &   12.724  &   11.833& \\
\enddata
\end{deluxetable}

\end{document}